\PassOptionsToPackage{pdfpagelabels=false}{hyperref} 

\documentclass[fleqn,usenatbib]{mnras}

\usepackage{newtxtext,newtxmath}
\usepackage{natbib}
\usepackage{graphicx}    % Including figure files
\usepackage{amsmath}    % Advanced maths commands
\usepackage[T1]{fontenc}
\usepackage{dblfloatfix} % To enable figures at the bottom of page
\usepackage{upgreek}   % Non italic greek letter
\usepackage{ae,aecompl}
\usepackage{siunitx}
\usepackage{tabularx}
\usepackage{lscape}
\usepackage{multirow}
\defcitealias{2003A&A...398..181V}{SL03}
\defcitealias{1998A&A...335.1029S}{S98}
\defcitealias{2013A&A...554A.104F}{FM13}
\defcitealias{1991A&A...244..205E}{E91}

\title[M1-67]{New insights into the WR nebula M1-67 with SITELLE}

\author[M. S\'evigny et al.]{Marcel S\'evigny,$^{1,3}$\thanks{E-mail: marcel.sevigny.1@ulaval.ca}
Nicole St-Louis,$^{2,3}$
Laurent Drissen,$^{1,3}$
and Thomas Martin$^{1,3}$
\\
$^{1}$D\'epartement de physique, de g\'enie physique et d'optique, Universit\'e Laval, Qu\'ebec (QC), G1V 0A6, Canada\\
$^{2}$D\'epartement de Physique, Universit\'e de Montr\'eal, C.P. 6128, Succ. Centre-Ville, Montr\'eal (QC), H3C 3J7, Canada\\
$^{3}$Centre de recherche en astrophysique du Qu\'ebec (CRAQ)\\
}

\date{Accepted XXX. Received YYY; in original form ZZZ}

\pubyear{2020}

\begin{document}
\label{firstpage}
\pagerange{\pageref{firstpage}--\pageref{lastpage}}
\maketitle

% Abstract of the paper
\begin{abstract}
We present a detailed study of M1-67, a well-known nebula around the population I Wolf-Rayet star WR 124 (WNh 8), based on datacubes obtained with the imaging Fourier transform spectrometer SITELLE at the Canada-France-Hawaii Telescope (CFHT). This allowed us to reconstruct detailed emission-line ratio maps that highlight clear orthogonal features from a chemical abundance point of view, a complete extinction map, as well as the electron density and temperature structures. In addition to this information, velocity maps were obtained shedding light on the bow shock structure due to the high velocity of WR124, qualified as a runaway star, which is about +190\,km\,s$^{-1}$ relative to the local ISM. Interaction between the latter structure and spherical and non-spherical outburst could explain the global morphology of M1-67. 
\end{abstract}

% Select between one and six entries from the list of approved keywords.
% Don't make up new ones.
\begin{keywords}
M1-67  -- imaging spectroscopy -- Wolf-Rayet Nebula -- Bubbles  -- jets and outflows
\end{keywords}

%%%%%%%%%%%%%%%%%%%%%%%%%%%%%%%%%%%%%%%%%%%%%%%%%%

%%%%%%%%%%%%%%%%% BODY OF PAPER %%%%%%%%%%%%%%%%%%

\section{Introduction} \label{section1}

M1-67 is a clumpy emission-line nebula surrounding the Population I Wolf-Rayet (WR) star WR 124 of spectral type WN8h \citep{2006A&A...457.1015H} that is known to be among the fastest runaway stars with a heliocentric radial velocity of $\sim$ 190.0 $\pm$ 7.4\,km\,s$^{-1}$ \citep{2007AN....328..889K}. The nebula is thought to be formed by material ejected by the star and interacting with the interstellar medium (ISM). 

First classified as an H{\sc ii} region \citep{1959ApJS....4..257S}, M1-67 was later thought to be a planetary nebula \citep{1964PASP...76..241B} and a ring nebula surrounding a WR star \citep{1975ApL....16..165C}. A more detailed study by \citet{1991A&A...244..205E} (hereafter E91) revealed a nitrogen enhancement and an oxygen deficiency, pointing to a reviewed classification as an ejected-type WR ring nebula. \cite{1998A&A...335.1029S} (hereafter S98) presented a kinematic study of the nebula based on a series of long-slit spectroscopic observations and found evidence for a spherical shell expanding at  46\,km\,s$^{-1}$ as well as signs of an  88\,km\,s$^{-1}$ bipolar outflow. More recently, \citet{2013A&A...554A.104F} (hereafter FM13) presented the first 2D study of the chemical abundance and kinematics of the central part and one external part of M1-67 with a spatial resolution of $\sim 3''$.

The WR phase is a late massive-star evolutionary stage characterized by high mass-loss rates and fast stellar winds \citep[e.g.][]{2007ARA&A..45..177C}. Studying the gas ejected by the star provides the opportunity to learn about its previous evolutionary phases. In the case of M1-67, most authors  (\citetalias{1998A&A...335.1029S}, \citealt{2003A&A...398..181V}, \citetalias{2013A&A...554A.104F}) concluded that the ejecta originated from one or two Luminous Blue Variable (LBV) outbursts. In this paper, we present results from our study of M1-67 with the imaging Fourier transform spectrometer (iFTS) SITELLE. Section 2 describes our observations and data reduction procedure and Section 3 presents our results as well as a discussion of our new findings in the context of what was previously known about this star and its nebula. Kinematics of the nebula are analyzed in Section 4 and our conclusions can be found in Section \ref{Conclusion}.

\section{Observations} \label{section2}

\begin{figure}
    \centering
    \includegraphics[width=1\columnwidth]{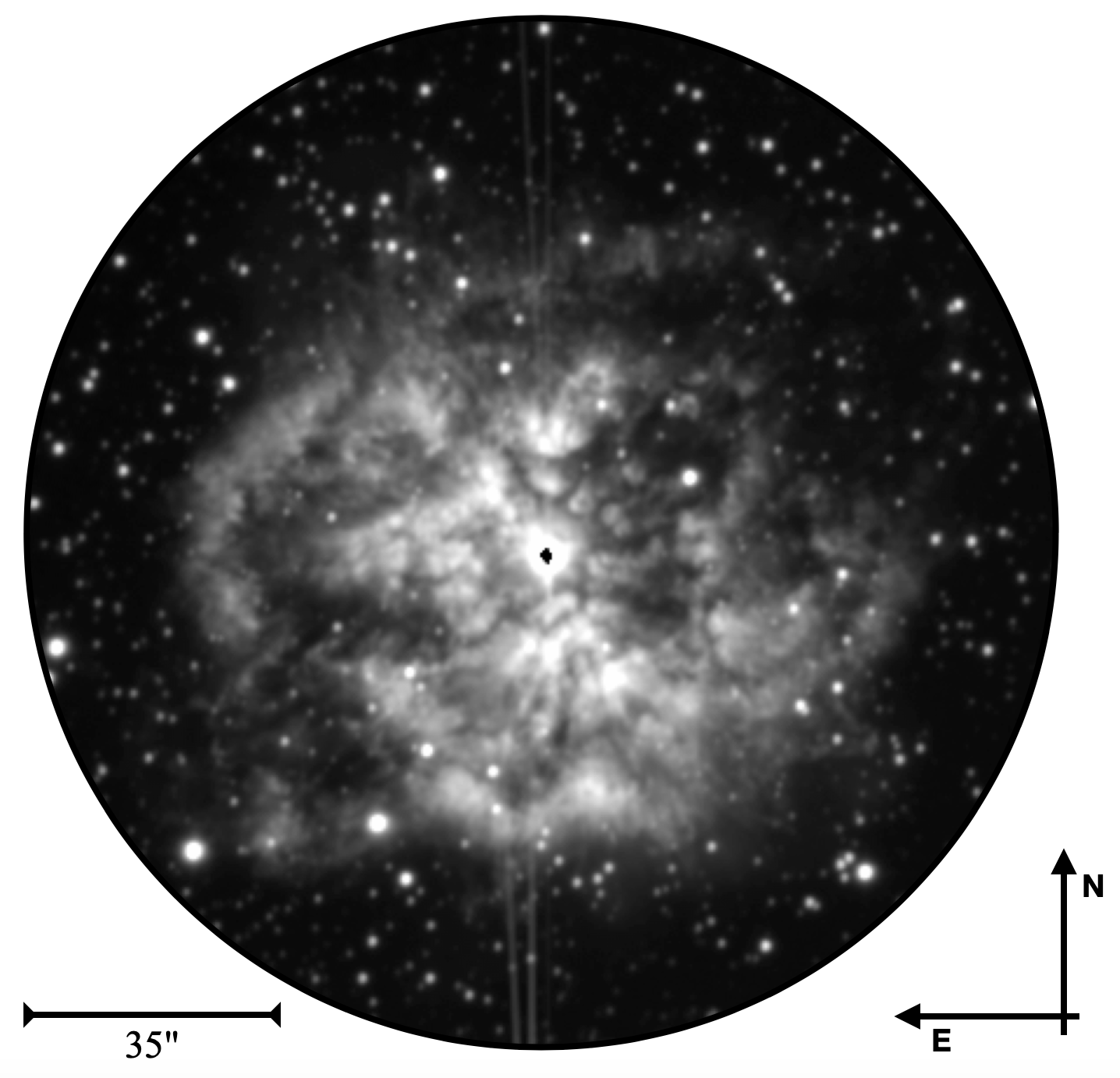}
    \caption{Flux-integrated SITELLE image of M1-67 in the SN3 filter [6470-6850 \si{\angstrom}] obtained by summing all interferograms in the raw datacube.}
    \label{fig1}
\end{figure}

We observed M1-67 with the iFTS SITELLE at the Canada-France-Hawaii Telescope (CFHT) in 2016 and 2017. SITELLE \citep{2019MNRAS.485.3930D} is perfectly adapted for large continuous-field spectroscopy since it provides spatially resolved spectra of sources in an $11' \times 11'$ field of view with a sampling of $0.32''$/pixel, in selected bandpasses of the visible range, with a spectral resolution adapted to the needs of the observer. Two e2v CCDs (2048 $\times$ 2064 pixels) record the interferograms, which are then Fourier transformed to produce photometrically and spectroscopically calibrated datacubes.
The instrument settings adopted for our various observations as well as the image quality are listed in Table \ref{table1-1}. The first datacube we obtained was with the SN3 filter and was secured during a science verification run with a full moon (the SN3 image integrated over all wavelengths is shown in Figure \ref{fig1}). The rest of the data was obtained during standard observing runs under dark sky conditions.

\begin{table*}
    \centering
        \caption{Instrumental settings used to obtain the SITELLE data of M1-67 according to the filter used}
    \begin{tabular}{lcccc} \hline \hline
        Filters & C2 &SN1 &SN2 & SN3 \\ \hline
        Observation date &July, 6, 2017 &July, 1, 2017 &July, 9, 2016  & May, 20, 2016  \\
        Spectral range [\si{\angstrom}] & 5590-6250& 3630-3860 &4820-5130 & 6470-6850   \\
        Number of steps & 226 &103 &136 & 264 \\
        Total exposure time [h] & 2.4 &1.4 &3.8 & 2.6 \\
        Spectral resolution R & 300 & 300 & 600 & 1300  \\
        Image quality [$\arcsec $] & 1.3 & 1.3 & 1.2 & 1.5  \\ 
        \hline
    \end{tabular}
    \label{table1-1}
\end{table*}

The SN3 datacube includes the strong H$\upalpha$ Balmer line as well as the [N{\sc ii}]$\uplambda$6548-84 and [S{\sc ii}]$\uplambda$6717-31 doublets (see Figure \ref{fig4}), the latter of which is often used to determine the electron number density. The SN2 datacube includes the H$\beta$ Balmer line that is crucial to determine the interstellar extinction when combined with another Balmer line, in this case, H$\alpha$. The main purpose of observing the nebula with the C2 filter was to detect the very faint [N{\sc ii}]$\uplambda$5755 line, with the goal of estimating the electron temperature when combined with the stronger nitrogen doublet in the SN3 datacube. Finally, the SN1 cube includes the [OII]$\uplambda$3727-29 doublet, for which the integrated flux can be used to estimate the oxygen abundance in specific regions of the nebula.

\subsection{Data Reduction} \label{section2.1}

The data reduction was carried out using ORBS, SITELLE's dedicated data processing and calibration software \citep{2014ascl.soft09007M}. The standard star used for the flux calibration was GD71 for all four datacubes. Well-established spectral reduction routines were applied to obtain the final data presented in this paper. For more details on the standard reduction, photometric calibration and other corrections using ORBS, see Sections 2 \& 3 of \citet{2018MNRAS.477.4152R} or Section 2 of  \citet{2018MNRAS.473.4130M}. 

\begin{table}
    \centering
        \caption{Comparaison of various fluxes and of the extinction coefficient of regions studied in previous papers. For each quantity, we present the ratio between their values and ours. Ratios that are not within the error interval are identified by a *.}
    \setlength{\tabcolsep}{7pt}    
    \renewcommand{\arraystretch}{1}
    \begin{tabular}{l|cc|ccc} \hline \hline
        Authors & \multicolumn{2}{c}{\text{\citetalias{1991A&A...244..205E}}} & \multicolumn{3}{c}{\text{\citetalias{2013A&A...554A.104F}}} \\ 
        Regions & A & C & 1 & 5 & 6 \\ \hline
         c$_{(\rm{H}\upbeta)}$ & 0.90 & 0.98 & 1.27* & 1.11* & 1.17* \\
         F(H$\upbeta$) & 1.06 & 0.99 & 0.95 & 0.99 & 1.01  \\
         F([O{\sc ii}]3727)  & 0.86 & / & / & / & / \\
         F([N{\sc ii}]6548) & 0.92 & 0.88 & 0.83* & 0.94 & 0.83* \\
         F(H$\upalpha$) & 0.91 & 0.90 & 0.88* & 0.86* & 0.88* \\
         F([N{\sc ii}]6584) & 0.89* & 0.87* & 0.86* & 0.96 & 0.89* \\
         F([S{\sc ii}]6717) & 1.07 & 0.92 & 0.93 & 1.10 & 0.85* \\
         F([S{\sc ii}]6731) & 1.08 & 1.02 & 0.91 & 1.04 & 0.80* \\
         %n$_{\rm e}$ & 1.48 & 2.18 & 1.21 & 1.18 & 1.01 \\
         %T$_{\rm e}$ & 0.89 & 0.97 & / & / & / \\
    \end{tabular}
    \label{table2.2}
\end{table}

Before combining information from different datacubes, it was also crucial to ensure that they were spatially aligned. To realign the SN2, SN1 and C2 observations with the SN3 datacube, we used the \textit{geotran} task in IRAF\footnote[1]{IRAF is distributed by the National Optical Astronomy Observatories (NOAO), which are operated by the Association of Universities for Research in Astronomy (AURA), Inc., under cooperative agreement with the National Science Foundation.}, which simply uses the position of background stars (which can be seen in the SN3's deepframe in Figure \ref{fig1}) and calculates a geometric transformation to apply to the other cubes. Finally, we used the OH lines within our datacubes to obtain an accurate wavelength calibration of our spectra. To complete the velocity correction, we applied the barycentric correction.

\section{Flux maps} \label{section3}

\subsection{Flux calibration and line profile fitting} \label{section3.1}

The data presented in this paper are among the first acquired by SITELLE in several bands (notably C2), and were reduced with the first version of the data calibration procedure for this instrument. Moreover, the SN3 cube was obtained during an engineering run with a Full Moon. Insuring that the flux calibration between datacubes is reliable is crucial to be able to combine emission lines from different filters to determine various physical parameters of the nebula. Therefore, we selected regions of the nebula previously studied by other authors to compare physical parameters such as the extinction coefficient and relative fluxes of strong lines that we obtained from our data with those determined by previous authors. 

ORCS, SITELLE's analysis software suite \citep{2020ascl.soft01009M}, is used to fit the observed line profiles for each pixel of the field of view (or an integrated region within it) and produce maps (or single values) of line intensities, radial velocity and velocity dispersion as well as their uncertainties. Because at each mirror step the flux from the entire waveband is acquired by the detectors, the photon noise from an FTS is distributed across all spectroscopic channels after the Fourier transform; this is different from a conventional spectrograph, for which the photon noise associated with a given spectroscopic channel depends on the photon count at the corresponding wavelength.This is taken into account by ORCS, which returns, for each parameter of the fit, the corresponding error. The S/N ratio is thus determined using these values.

In Table \ref{table2.2}, we present the ratio between our measured values of the extinction coefficient and relative fluxes in specific lines and that of previous authors. To carry-out this comparison, we selected the same locations and shapes (simple rectangles for \citetalias{1991A&A...244..205E} and more complex shapes for \citetalias{2013A&A...554A.104F}) for different regions studied by previous authors.  Because there is still somewhat of an uncertainty on the positions of the various regions, we compared our H$\upbeta$ flux integrated over the selected regions with that from previous studies. We then slightly adjusted the position until we found the flux to be the same. We then considered that location as the correct one. Since the atomic data used for the determination of the electronic density and temperature might be different in both studies, we decided not to compare those physical quantities. From this table, it can be seen that our data are compatible (within the errors) with those obtained in earlier studies (e.g. \citetalias{1991A&A...244..205E} and \citetalias{2013A&A...554A.104F}). Our determination of the extinction coefficient agrees very well with that of \citetalias{1991A&A...244..205E} whereas that of \citetalias{2013A&A...554A.104F} is higher by 10-30\%. Our H$\beta$ flux, on the other hand, agrees well with all previous measurements.  The line fluxes we measure in the lines listed in Table \ref{table2.2} agree well with those of \citetalias{1991A&A...244..205E} with the exception of the [N{\sc ii}]$\lambda$6584 line for which we measure a value that is $\sim$10\%\ higher.  We have more disagreements with the values measured by \citetalias{2013A&A...554A.104F} with differences reaching between 11-20\%. We also compared the H$\upalpha$ flux for the entire nebula with data obtained with the HST (\citetalias{1998A&A...335.1029S}). These authors measured a total reddening corrected flux of 2.08 $\pm$ 0.12 $\times$ 10$^{-10}$ ergs$^{-2}$ s$^{-1}$ using the reddening coefficient determined by \citetalias{1991A&A...244..205E}. Using the same method and the same correction factor, we obtained a total flux of 1.97 $\times$ 10$^{-10}$ ergs$^{-2}$ s$^{-1}$ which is in excellent agreement with the HST data.

\begin{figure}
    \centering
    \includegraphics[width=\columnwidth]{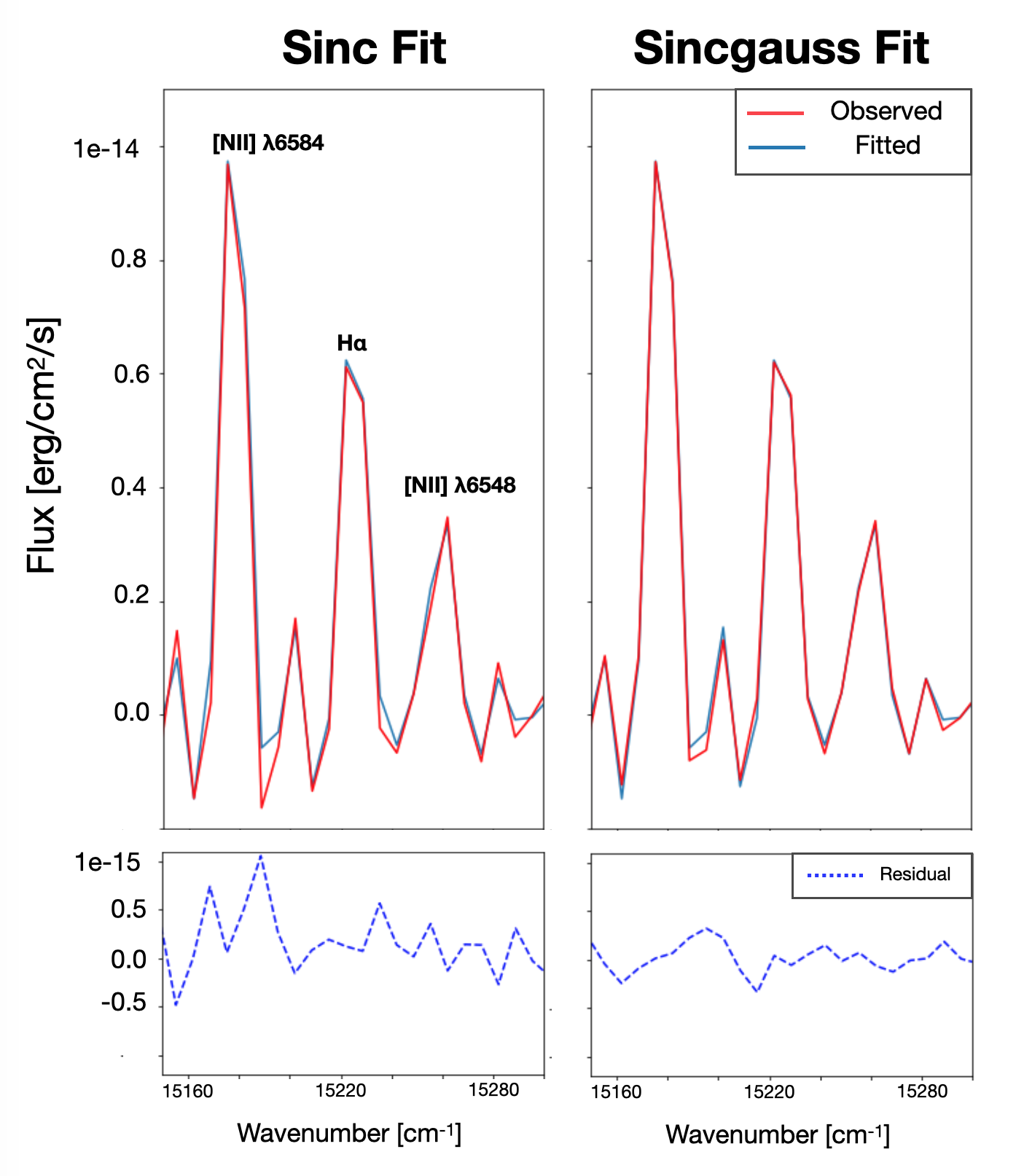}
    \caption{Line fits using two different functions with the ORCS software. The left and right panels show fits of the same region, but with a different fitting function, respectively sinc and sincgauss function. The fits are in red and the observations in blue. The dashed blue lines in the bottom panels are the residuals between the fits and the observation highlighting the best fitting function.}
    \label{fig2}
\end{figure}

SITELLE's instrument line shape (ILS) is a sinc function. However, any line broadening caused by turbulent motion or a velocity gradient along the line of sight will transform the natural ILS into a so-called sincgauss function: the convolution of a sinc and a gaussian  \citep{2016MNRAS.463.4223M}.
Figure \ref{fig2} compares fits using these two line shapes for the same spectrum of a specific region. This figure shows that, while the sinc function is a good solution, a better fit is obtained with a sincgauss.

\begin{figure}
    \centering
    \includegraphics[width=\columnwidth]{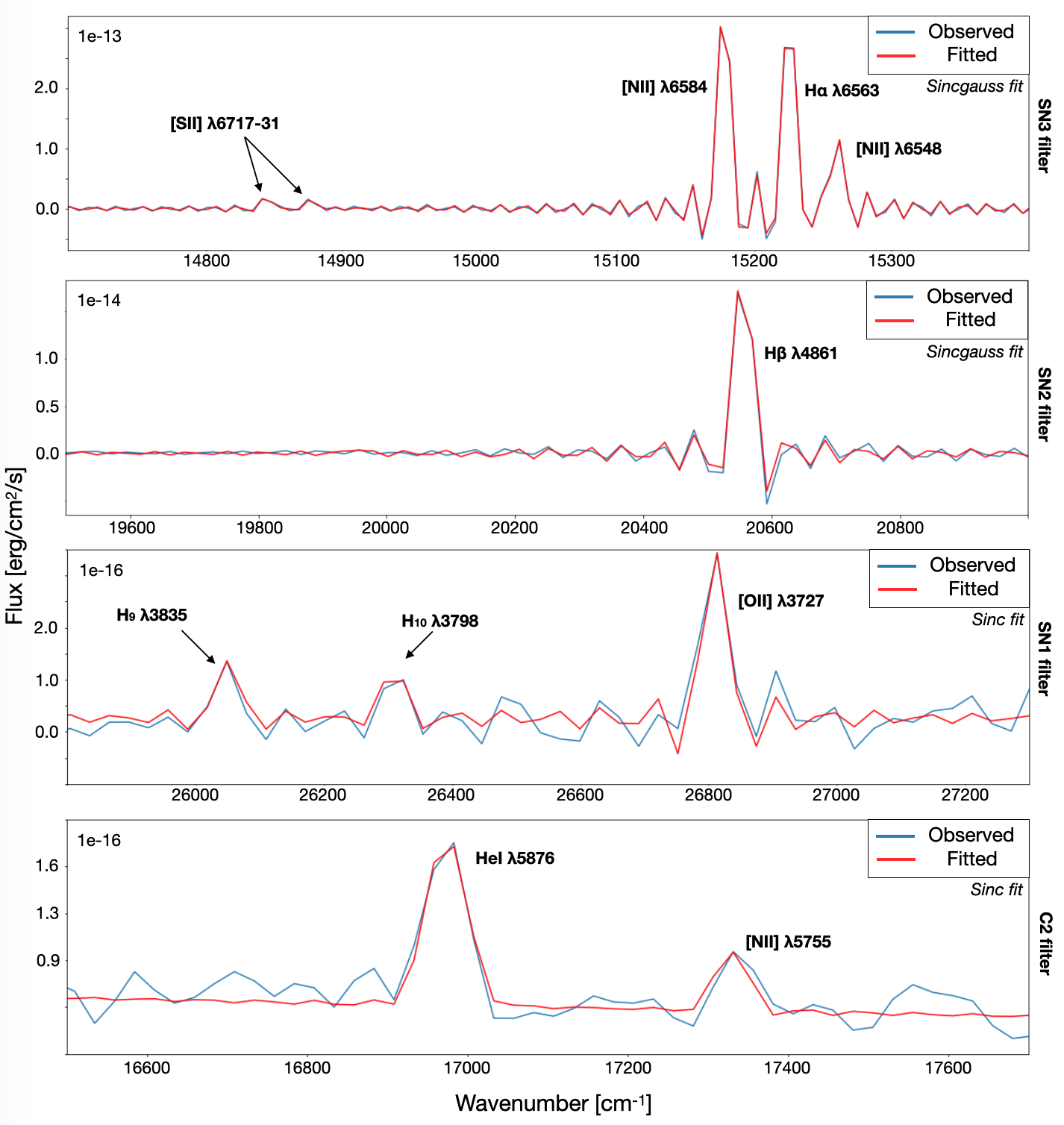}
    \caption{Spectrum of M1-67 for Region 3 (corresponding to the R3 region shown in Figure \ref{fig4}) for the SN3, SN2, SN1 and C2 filters respectively. The blue line is the raw spectrum observed and the red line is the best fit obtained with ORCS with a sincgauss or a sinc profile depending on the spectral resolution obtained with the filter.}
    \label{fig3}
\end{figure}

\begin{figure}
    \centering
    \includegraphics[width=\columnwidth]{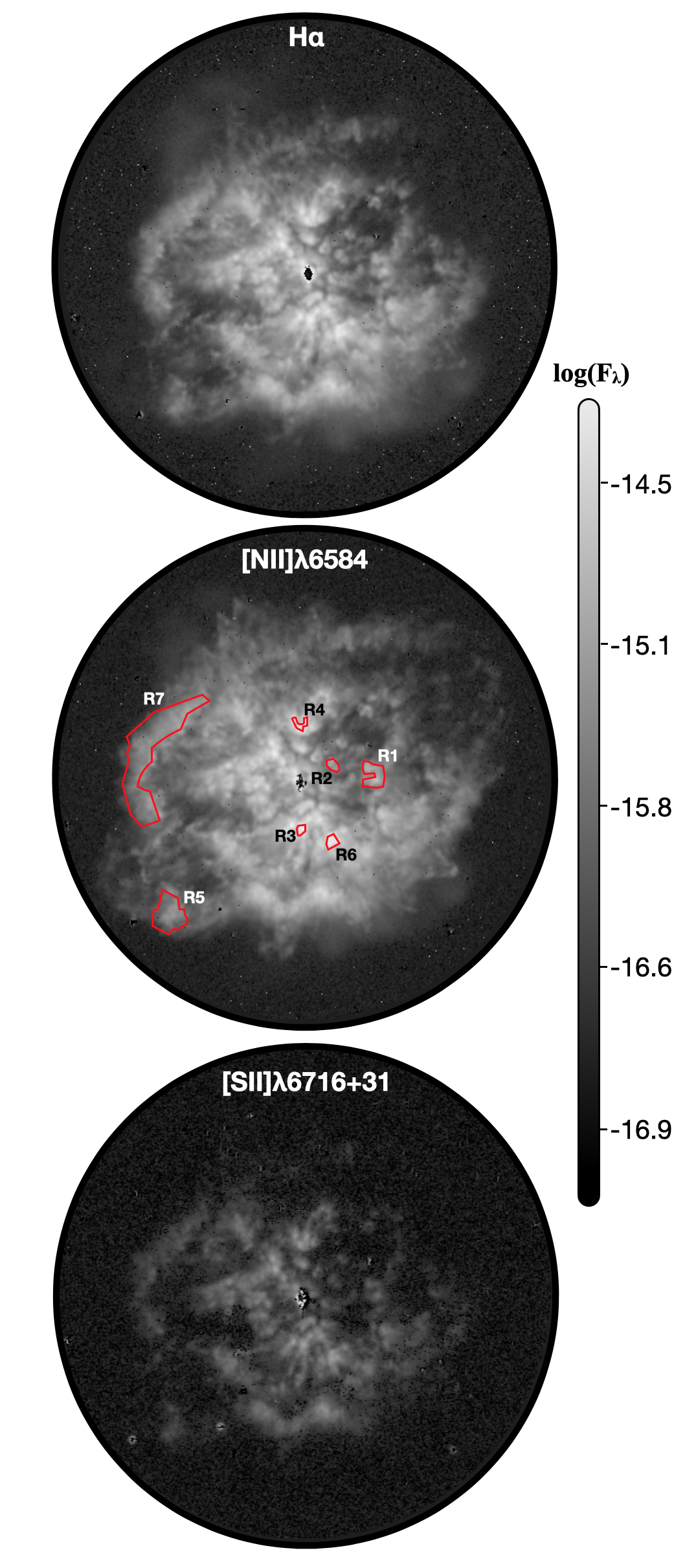}
    \caption{Flux maps in the H$\upalpha$ line and the [N{\sc ii}]$\uplambda$6584 and [S{\sc ii}]$\uplambda$6717+31 doublets from the SN3 datacube in units of erg cm$^{-2}$ s$^{-1}$ (as for all other flux maps included in this paper). For the H$\upalpha$ and [N{\sc ii}]$\lambda$6584 maps, we have retained only pixels with S/N $\ge$ 3 while for the map for the [S{\sc ii}] doublet in the bottom panel, we have kept all pixels. In the middle map ([N{\sc ii}]$\uplambda$6584), we indicate the seven flux-integrated regions specifically studied in this paper in red contours (see Table \ref{table2.4} and Table \ref{table2.5}).}
    \label{fig4}
\end{figure}

Therefore, we fitted the various emission lines with a single component sincgauss for the SN3 filter since the spectral resolution for this filter was sufficiently high. For the other three filters (SN2, SN1 and C2), a single component sinc function was used. Example fits are shown in Figure \ref{fig3}. For the H$\upalpha$ and [N{\sc ii}] lines, the resulting flux maps are presented in the first two panels of Figure \ref{fig4}. For these two relatively strong lines, we retained only pixels with a signal-to-noise ratio (S/N) $\ge$ 3. However, for the much fainter [S{\sc ii}] doublet shown in the bottom panel of this figure, all pixels were retained. 

\begin{figure}
    \centering
    \includegraphics[width=\columnwidth]{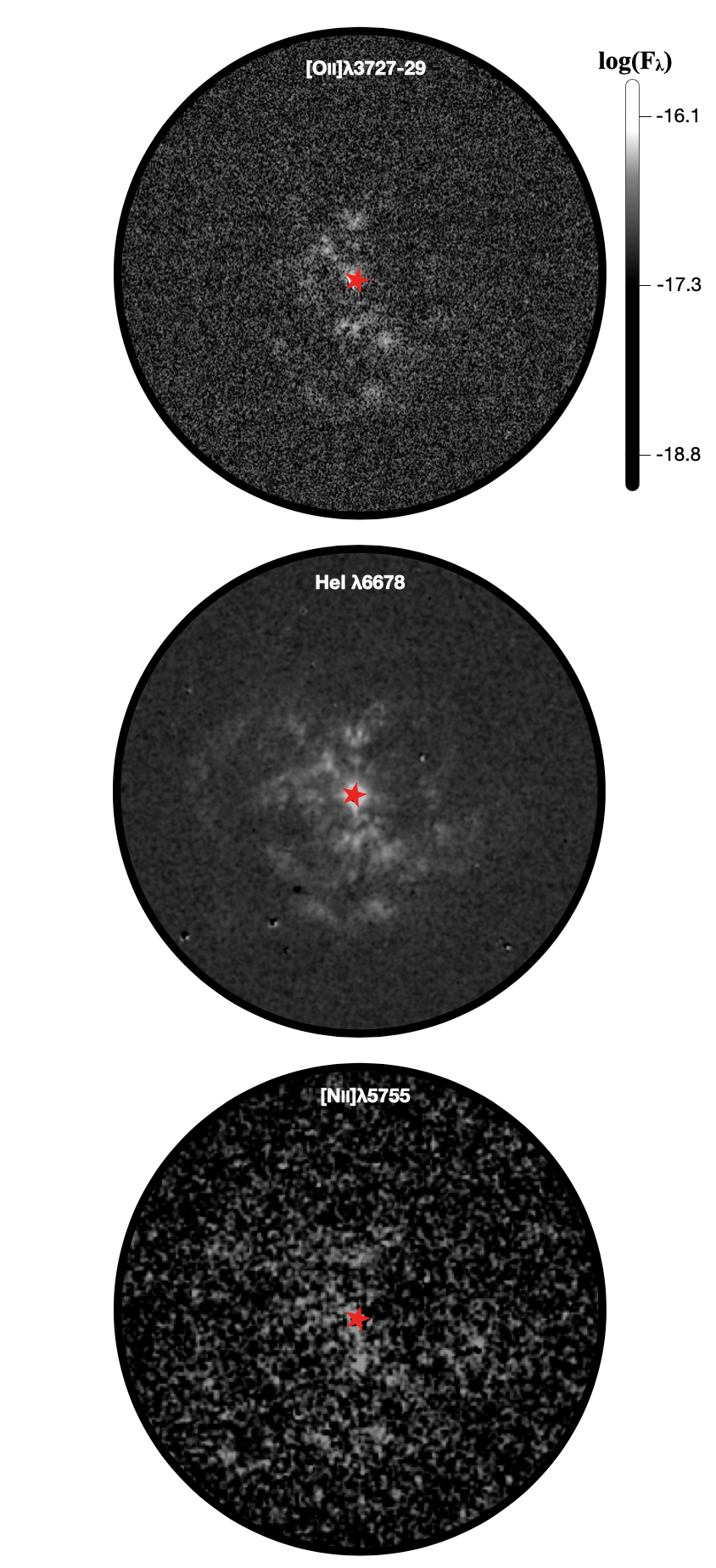}
    \caption{Flux or intensity maps of M1-67 for fainter lines from the SN1, SN3 and C2 datacubes. The top panel presents a flux map for the [O{\sc ii}]$\uplambda$3727-29 doublet and the middle and bottom panels are gaussian-convolved, continuum-subtracted images of the He{\sc i}$\uplambda$6678 and the [N{\sc ii}]$\uplambda$5755 lines.}
    \label{fig5}
\end{figure}

Our observations also allowed us to obtain 2D images of M1-67 in very faint lines. These are shown in Figure \ref{fig5}. For the [OII]$\uplambda$3727-29 doublet, we show a complete flux map. For the even fainter HeI$\uplambda$6678 and [N{\sc ii}]$\uplambda$5755 lines, it was impossible to obtain a pixel-to-pixel flux map. Instead, we show images directly extracted from the data cube at the maximum of the emission line (from which the adjacent continuum has been subtracted); these images, convolved with a 2-pixel gaussian kernel, reveal the regions where the lines are detected. We note that the central star is very bright in the HeI$\uplambda$6678 image since this line is present in emission in its spectrum. 

\begin{figure}
    \centering
    \includegraphics[width=\columnwidth]{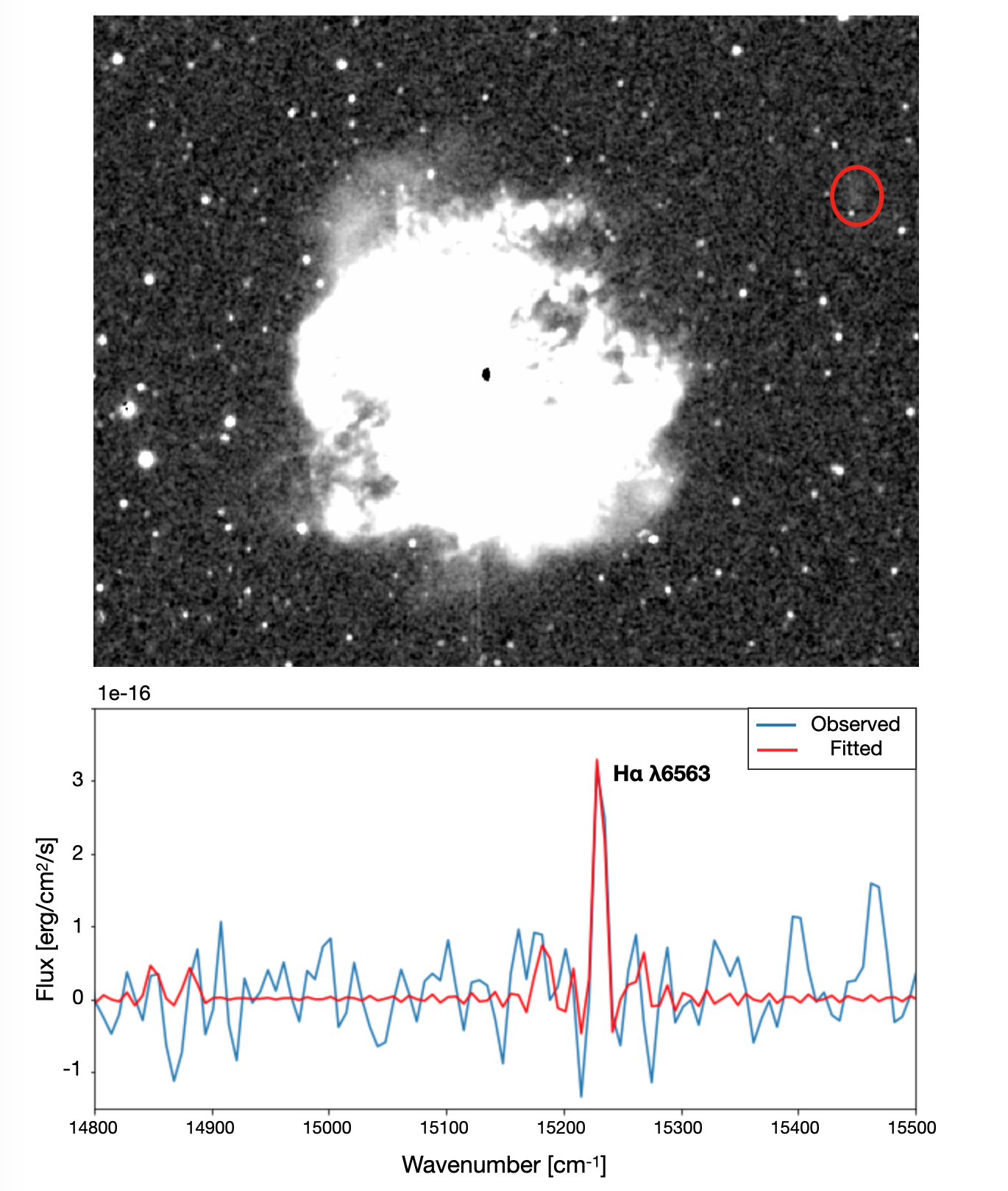}
    \caption{(Upper panel) H$\upalpha$ image of M1-67 showing the brightest of several faint structures away from the main body of M1-67, centered at 19h11m24s, +16$^o$52$'$25$"$. This image is the average of two adjacent channels of the data cube in which the structure stands out, convolved with a 0.5$''$ gaussian kernel to increase the contrast between the faint structure and the background. (Lower panel) Spectrum of the faint structure. Note the fact that the [NII] lines are much weaker (lost in the noise) than H$\upalpha$, contrary to the main body of M1-67.}
    \label{fig15}
\end{figure}

The various maps presented here all show the well-known filamentary structure of the nebula elongated in the NW-SE direction. We also note a more diffuse extension, particularly strong in H$\upalpha$, orthogonal to the main elongated body. Such a structure was also detected by \citetalias{2013A&A...554A.104F} in the optical and \citet{2018ApJ...869L..11T} in the infrared (WISE 12$\upmu$m, Spitzer 24$\upmu$m and Herschel 70$\upmu$m images). We also detect hints of very faint structures, visible in H$\upalpha$ only, around the main nebula. The most obvious is located at 19h11m24s, +16$
^o$52'25'' (outside of the field of view of Figure \ref{fig4}, but shown in the upper panel of Figure \ref{fig15}).  Its spectrum is displayed in the bottom panel and clearly shows the H$\upalpha$ line at a velocity of +91 $\pm$ 12 km/s, without the presence of [NII] above the noise. It is tempting to suggest that this feature, if indeed it is physically associated with M1-67, originates from a previous evolutionary phase; its radial velocity is indeed similar to that of one of the velocity components of the main nebula (see Figure \ref{fig12}), but deeper data are necessary to confirm this association.

\subsection{Reddening correction maps} \label{section3.2}

We obtained the first complete extinction coefficient map, $c_{\text{H}\upbeta}$, from the H$\upalpha$/H$\upbeta$ ratio using :

\begin{equation}
    c_{\text{H}\upbeta} = 2.78 \times \log_{10}\left(\frac{ F'(\text{H}\upalpha)/F'(\text{H}\upbeta)}{3.05}\right)
\end{equation}

\medskip

where the 3.05 factor is the H$\upalpha$/H$\upbeta$ Balmer decrement for Case B in the low-density limit at a temperature of 5 000 K (\citealt{2006agna.book.....O}), according to previous temperature determinations by \citetalias{1991A&A...244..205E}. F$'$(H$\upalpha$) and F$'$(H$\upbeta$) are the observed reddened H$\upalpha$ and H$\upbeta$ integrated line fluxes from our SITELLE data. 
The factor 2.78 was obtained using values of the reddening function for a standard extinction curve (\citealt{2006agna.book.....O}). The extinction coefficient can then be used to correct for dust absorption following the method suggested by \cite{2010ApJ...711..619M}.  The dereddened intensity at a specific wavelength, I($\uplambda$), is given by the following expression where f($\uplambda$) is the reddening function:

\begin{equation}
    \frac{I(\uplambda)}{I(\text{H}\upbeta)} = \frac{F'(\uplambda)}{F'(\text{H}\upbeta)} \times 10^{c_{\text{H}\upbeta} f(\uplambda)}
\end{equation}

\begin{figure}
    \includegraphics[width=\columnwidth]{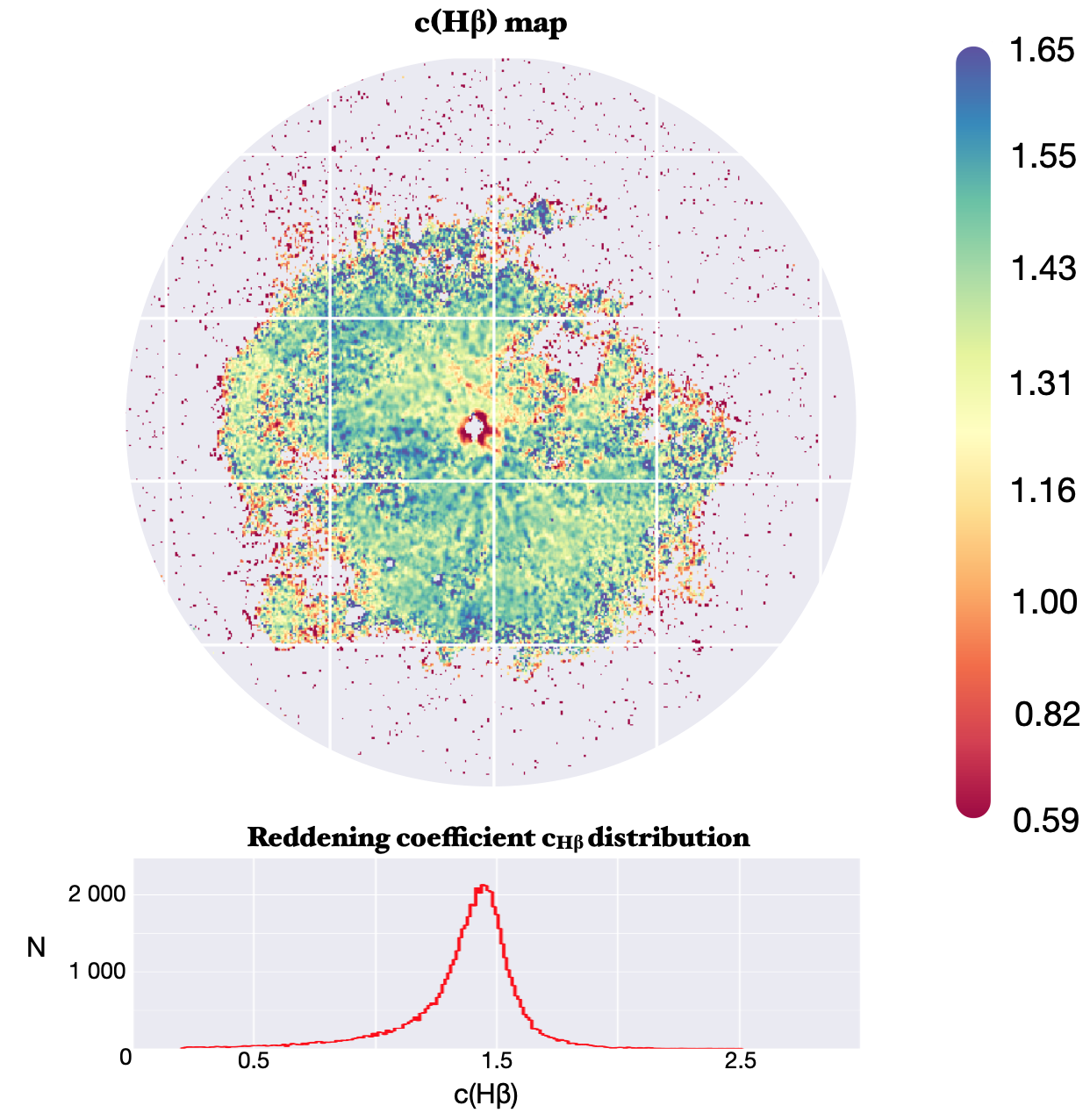}
    \caption{ Extinction coefficient map, $c_{\text{H}\upbeta}$, for M1-67. Bottom panel: Histogram of the extinction coefficient $c_{\text{H}\upbeta}$ determined from the H$\upalpha$, contrary to
the main body of M1-67/H$\upbeta$ maps.}
    \label{fig6}
\end{figure}

In the top panel of Figure \ref{fig6}, we present our extinction coefficient map, $c_{\text{H}\upbeta}$, for pixels with a S/N above 3, while the bottom panel presents the histogram of the extinction coefficient values for the M1-67 nebula obtained using the ratio map. This histogram can be compared with the one presented by \citetalias{2013A&A...554A.104F}. We find a median value for $c_{\text{H}\upbeta}$ of $1.41 \pm 0.05$ and a FWHM of 0.1, while they find a mean value of $1.85\pm0.10$ for the central regions and $2.11\pm0.08$ for the outer regions. Although our histogram has a similar shape to theirs, our values seem to be systematically lower. In the region corresponding to their "outer region" we find values closer to the range $c_{\text{H}\upbeta}=1.4-1.7$ instead of 2.1, while the lower values in the central regions are in the range 1.2-1.4 instead of 1.8 (see Table \ref{table2.2}). We can also compare our values to those of \citetalias{1991A&A...244..205E}. For four different slit positions, these authors find a uniform value of $c_{\text{H}\upbeta}=1.35$ with an error of $\pm 0.15$, which is in good agreement to our mean value of $1.31 \pm 0.05$. Our detailed reddening coefficient map $c_{\text{H}\upbeta}$ shows an inhomogeneous structure with filaments that seem to be oriented mainly radially (similar radial features are also observed in further ratio maps, see Section 3.3). Reddening is associated with the presence of dust, which was indeed detected with infrared images and spectroscopy by \cite{2016A&A...588A..92V} using Herschel. These authors found that the dust appeared clumpy and mixed with the ionized gas. Unfortunately, their maps have a spatial resolution that is too low to determine if our regions of higher extinction correspond to regions where dust is located. The WISE and Spitzer infrared images presented by  \cite{2018ApJ...869L..11T} are also of insufficient spatial resolution to be able to compare in detail with the extinction features we find with our SITELLE data. 

We used our extinction coefficient map to correct our flux maps for reddening. The general appearance of all maps was not significantly modified by these corrections. Therefore, to avoid losing information from fainter external regions, we choose to only present non-dereddened maps. However, all physical parameters we determined are based on reddening-corrected fluxes, including the relative fluxes.

\subsection{Ratio maps} \label{section3.3}

As mentioned above, for the final H$\upalpha$ and [N{\sc ii}] line maps presented here, we retained only pixels with a S/N above 3. Note, however, that although the majority of pixels in our maps have a S/N of 10 and above, we chose to keep the rejection criterion low to allow the analysis of weak external regions of the nebula. By comparing the H$\upalpha$ and the [N{\sc ii}] maps presented in Figure \ref{fig4}, a structure becomes apparent in the direction perpendicular to the more well-known structure of M1-67. Indeed while the [N{\sc ii}]$\uplambda$6584 flux extends further out than the H$\upalpha$ flux in the NW-SE direction, perpendicular to this direction (NE-SW), the H$\upalpha$ flux reaches a more external region than the [N{\sc ii}]$\uplambda$6584 emission.

\begin{figure}
    \centering
    \includegraphics[width=1.1\columnwidth]{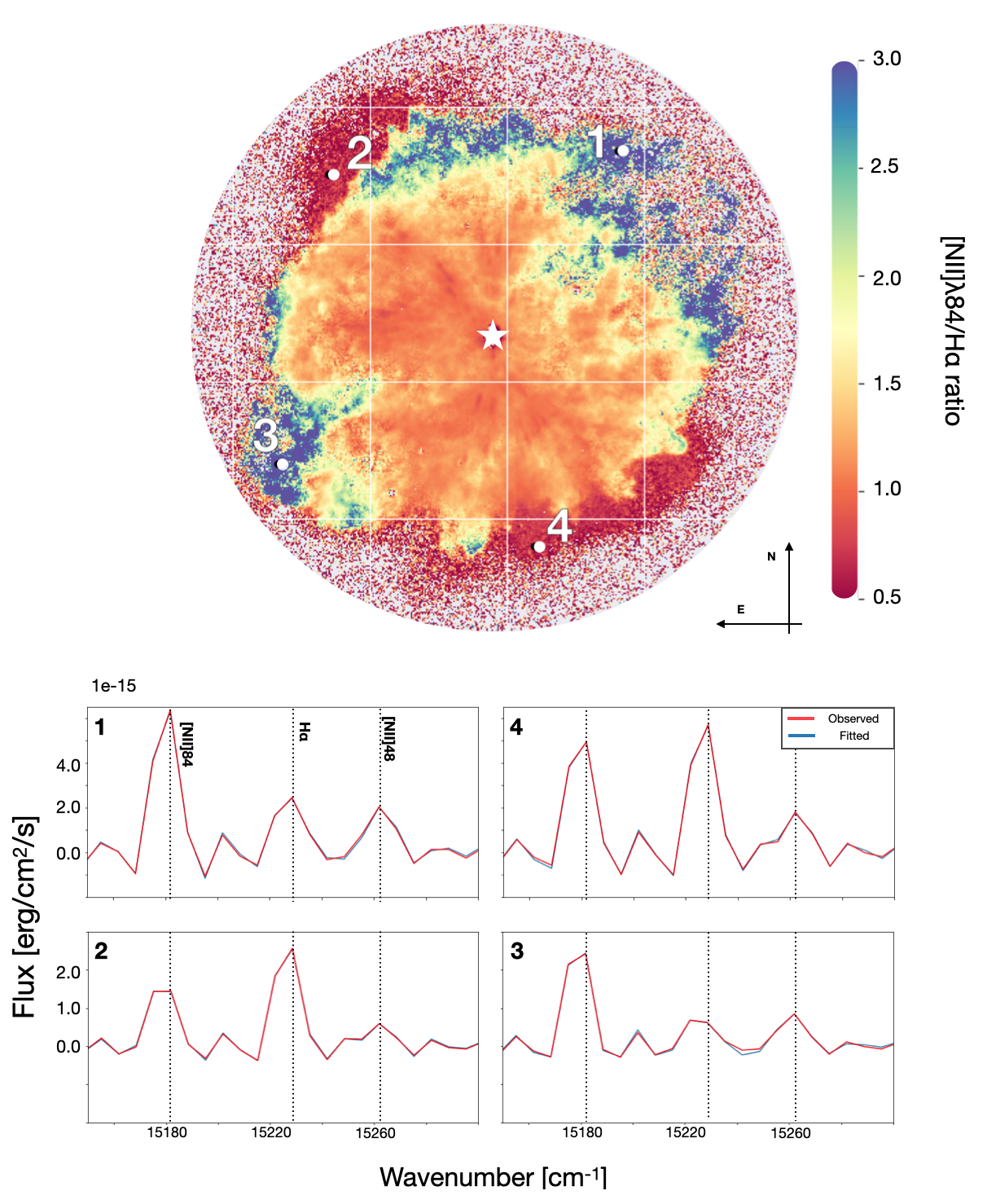}
    \caption{Top: [N{\sc ii}]$\uplambda$6584 to H$\upalpha$ line ratio map where red regions represent a stronger hydrogen line compared to nitrogen, while blue regions correspond regions where the hydrogen line is weaker.  Note that the entire nebula is enriched in nitrogen, as shown in the 4 spectra in the lower panels. We chose different regions of the nebula to show their symmetry on the two axes and the behaviour of the nitrogen and hydrogen lines.}
    \label{fig7}
\end{figure}

This orthogonality is further emphasized in Figure \ref{fig7}, where we present a colour-coded line-ratio map of [N{\sc ii}]$\uplambda$6584 to H$\upalpha$. In the NE-SW direction, the regions coloured in red show zones where the [N{\sc ii}] lines are fainter than H$\alpha$ (log F$'$[N{\sc ii}] / F$'$H$\alpha$) $<$ 0) while in the NW-SE direction, they are stronger (log (F$'$[N{\sc ii}] / F$'$H$\alpha$) $>$ 0), reaching values up to 3 times higher. Note that the flux measured in the [N{\sc ii}]$\uplambda$6548-84 doublet remains relatively high everywhere indicating that the whole nebula is indeed nitrogen enhancement. It seems to be the flux of the H$\upalpha$ line that is more variable. The spectra plotted in the four panels below the ratio map have been obtained for specific regions identified by white circles on the panel above.  As can be seen, hydrogen is only stronger than [N{\sc ii}] along the NE-SW axis (regions 2 and 4).

\begin{figure}
    \centering
    \includegraphics[width=1.1\columnwidth]{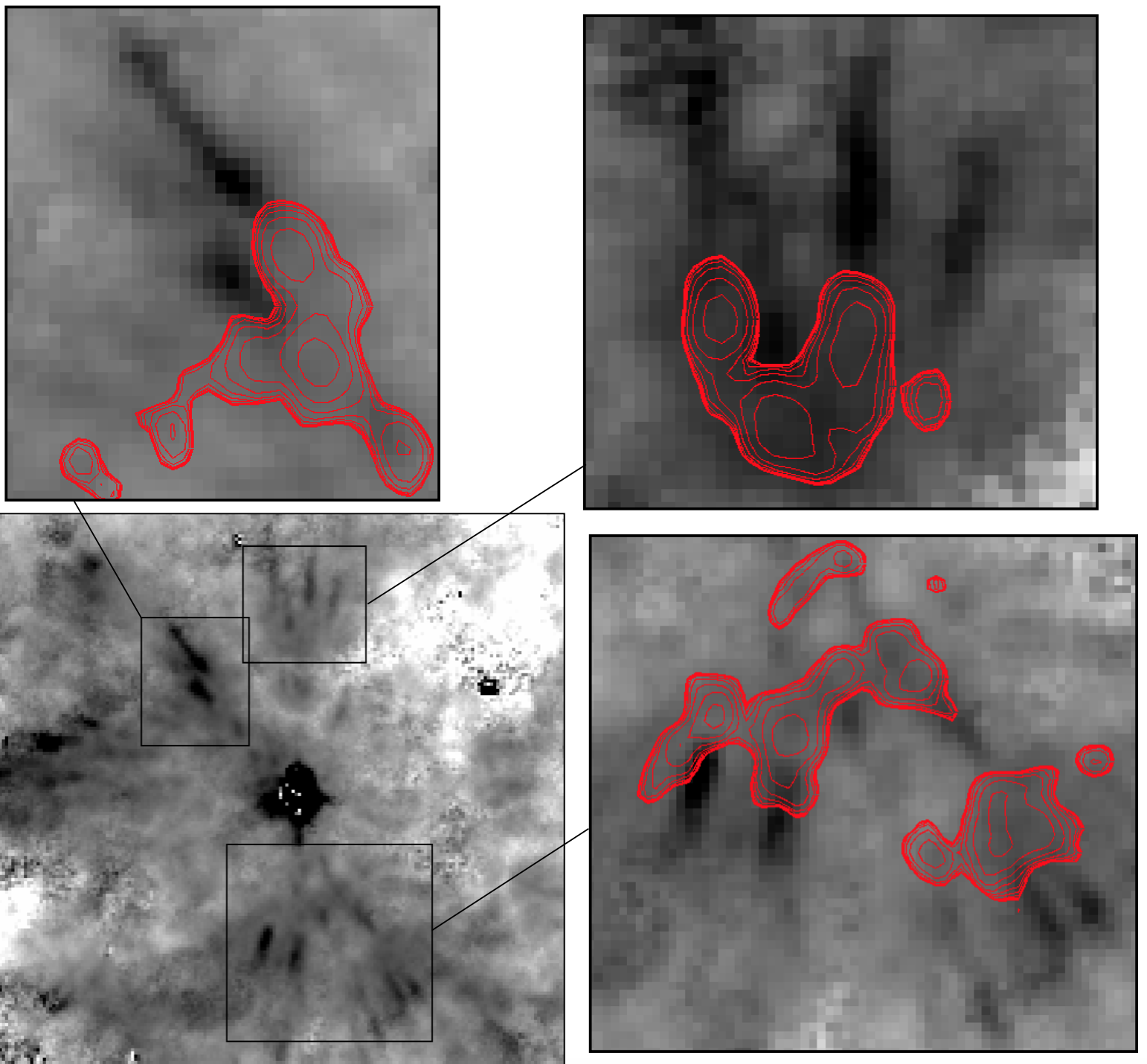}
    \caption{Central part of the [N{\sc ii}]$\uplambda$6584 to H$\upalpha$ line ratio map emphasizing the thin radial features. In the enlargements of the most outstanding of them, red contours of the H$\upalpha$ flux are superimposed on the line ratio.}
    \label{fig7b}
\end{figure}

There are two possibilities to explain the observed nitrogen/hydrogen line ratio behaviour highlighted in the top panel of Figure \ref{fig7} and in the four spectra presented below the ratio map. The first is a difference in N enrichment relative to hydrogen in the NW-SE and NE-SW directions, caused by an anisotropic ejection from the current or a past evolutionary phase, reminiscent of the peculiar shape of the $\upeta$ Car nebula, for example. This would imply that the ejection in the NW-SE would be more enriched in N than the one in the NE-SW direction. It could also be created from an anisotropic wind \citep{1983A&A...120..113M}. Such a wind would require a high rotation velocity. The second possibility is that it is caused by a major temperature and/or density differences between the two regions, affecting the photoionization state of the gas. To distinguish between these different scenarios, we need to construct temperature and electron density maps. 

We note in Figure \ref{fig7} the presence of a dozen thin (1$''$ wide) filamentary structures with a lower than average [NII]/H$\alpha$ ratio ($\sim$0.85; their immediate surroundings being at $\sim$1.05) extending radially out from the central star. Figure \ref{fig7b} shows an enlargement of the central regions of the nebula, highlighting these intriguing features. It is worth noting that they do not have a counterpart in either the H$\alpha$ (from SITELLE of Hubble Space Telescope) nor the [NII] flux images; they only show up in the line ratio map. Interestingly however, superimposing the line ratio and H$\alpha$ maps reveals that the radial structures originate from the opposite side (as seen from the star) of bright, dense H$\alpha$ knots. Since these features are not visible in the individual images, they cannot represent material ejected like bullets from the star. We do not have an obvious explanation for these features, but they could represent regions in the nebula shadowed from the star's direct UV flux, or elongated vortices resulting from the current stellar wind sweeping through dense knots previously ejected. Note that we find no clear link with the radial structures that appear in the extinction coefficient map shown in Figure \ref{fig6}.

\subsection{Temperature and electron density} \label{section3.4}

To obtain the temperature (T$_{\rm e}$) and electron density (n$_{\rm e}$) in the nebula requires the flux of faint lines, such as the [S{\sc ii}]$\uplambda$6717-31 doublet and the [N{\sc ii}]$\uplambda$5755 line. 
  
The ratio between the two components of the sulphur doublet is a good first approximation of the electronic density (up to n$_{\rm e}= 10^5$ \, cm$^{-3}$) in spite of a weak dependance on the temperature  (\citealt{2006agna.book.....O}). However, to obtain a more accurate estimate, we used the PyNeb library, that solves the equilibrium equations for an n-level atom and works iteratively with the ratio between the [S{\sc ii}]$\uplambda$6731 and $\uplambda$6717 lines as well as that of the [N{\sc ii}]$\uplambda$5755 and $\uplambda$6584 lines as input (see \citealt{2015A&A...573A..42L} for more details on this method). 

\begin{figure}
    \centering
    \includegraphics[width=1.1\columnwidth]{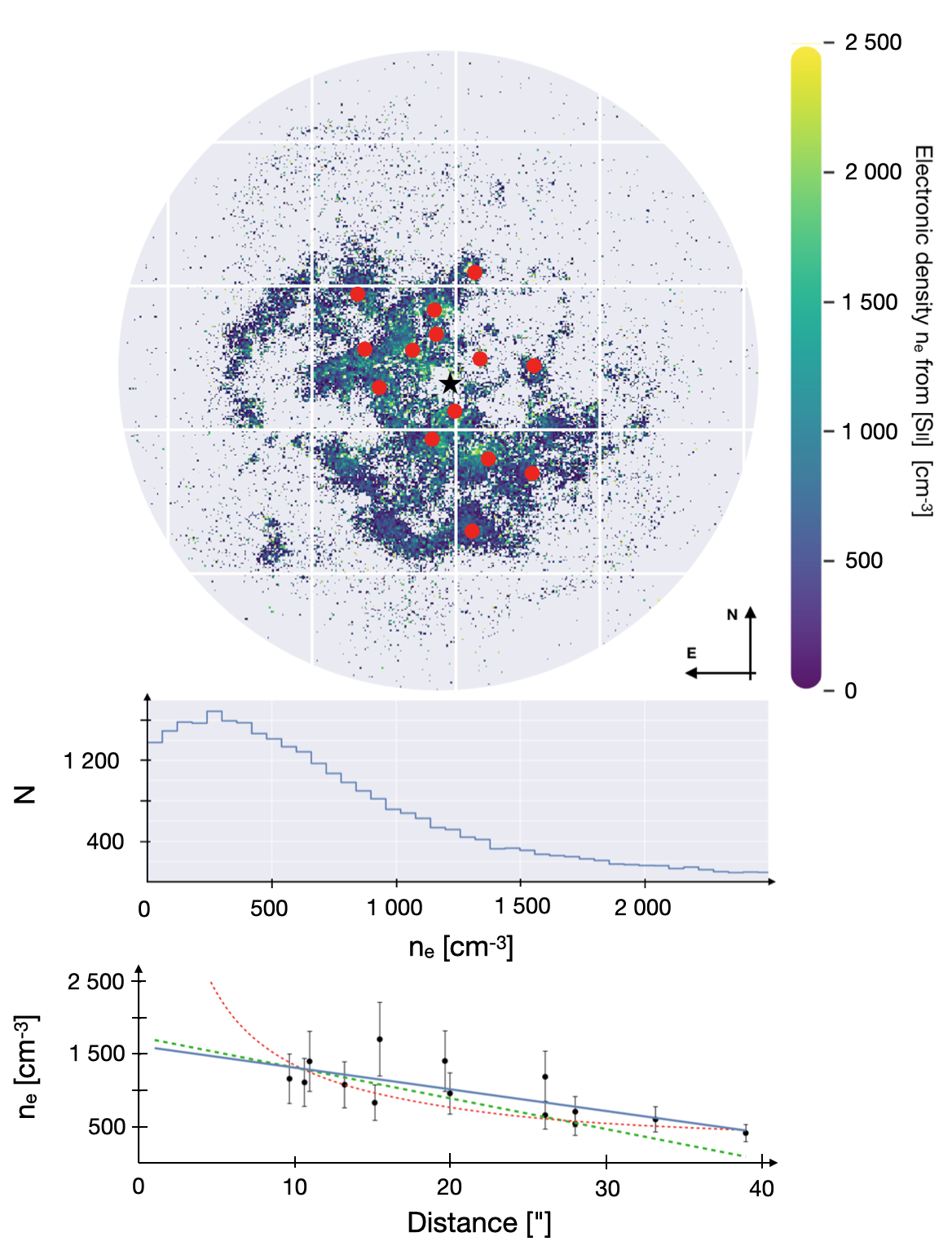}
    \caption{Top panel : Electron density map obtained from the ratio between the [S{\sc ii}]$\uplambda$6717 and $\uplambda$6731 lines illustrating the decreasing electron density with distance from the central star.  In the middle panel, we plot the histogram of the electron density values for the entire map showing that the distribution peaks around 300 cm$^{-3}$. In the bottom panel, we show the electron density measured over small regions indicated by red dots on the top map and plot their values as a function of distance from the star. The solid blue line shows a linear decrease with a slope of -29.97 cm$^{-3}$/$"$ and the red dashed line a power law with an exponent of -0.8. The green dashed line represents the linear decrease (least-square fits) obtained by \citetalias{2013A&A...554A.104F}.}
    \label{fig8}
\end{figure}

Because of the faintness of these lines, we were not able to determine the electron density for many of the external regions. In the top panel of Figure \ref{fig8} we present our $\rm n_e$ map for regions in which the flux is detected in the [S{\sc ii}] lines. One can see that overall, the densest (yellow-green) regions, $\rm n_e \sim 2 000$ cm$^{-3}$, are closer to the central star, while the most external regions display a modest density of $\sim 500$ cm$^{-3}$. A histogram of the density values for all pixels visible in the density map is presented in the middle panel of this figure. We find a broad distribution peaking around 300 cm$
^{-3}$ with a shallow tail reaching values as high as $\sim 2\,000$ cm$^{-3}$. Our distribution can be compared to that of \citetalias{2013A&A...554A.104F} who found a distribution peaking at $\sim 800$ cm$^{-3}$ in the central regions and $\sim 500$ cm$^{-3}$ in the only outer regions they observed. These values are systematically higher than the ones we found. \citetalias{2013A&A...554A.104F} also find a linearly decreasing electron density distribution which they claim is mainly in the NE-SW direction. We verified that the differences in the spatial sampling had a negligible effect on the values of n$_e$ and therefore could not explain the differences. On the other hand, such a discrepancy with \citetalias{2013A&A...554A.104F} could possibly be explained by the fact that they used a different atomic database (incorporated in the TEMDEN task in IRAF) from the one we used.

We also investigated the behaviour of $\rm n_e$ as a function of the distance from the star ($r$). We selected small regions distributed over the entire nebula (identified by red dots on our $\rm n_e$ map) and plot the value of n$_e$ in these regions as a function of $r$ in the bottom panel of Figure \ref{fig8}. We find what appears to be a linear decrease of the electron density with the distance from the star with a slope of -29.97 cm$^{-3}$/$"$ and a Spearman's Rank coefficient of 0.7. \citetalias{1998A&A...335.1029S} also quantified the behaviour of $\rm n_e$ as a function of distance from the star and found a power law relation with an index of -0.8, when assuming a cutoff in the nebula at $\sim 50"$. For comparison, we plot such a power law as a red dotted line in the bottom panel of Figure \ref{fig8} and we also add the linear decrease found by \citetalias{2013A&A...554A.104F} as a dashed green line. The error bars on our estimates are too large to be able to distinguish between either a linear or power-law decrease with distance. However, we can conclude that the decreasing electron density with distance from the central star is not limited to the NE-SW direction but is more radial in nature. Such a behaviour is compatible with an ejection in a pre-WR evolutionary phase such as a LBV or RSG.

The electron temperature is much more difficult to measure as the [N{\sc ii}]$\uplambda$5755 line is even weaker than the sulphur doublet. As for the electron density, we selected regions of the nebula and integrated the flux within them until the S/N was sufficiently high (S/N $\sim 3$). Our selected regions are indicated in the middle panel of Figure \ref{fig4} and labelled R1 through R7. These regions were selected based on their [N{\sc ii}]$\uplambda$5755 flux, but also to investigate knots at different distances from the central star in order to obtain a global view of the behaviour of the temperature in the nebula. External regions where the [N{\sc ii}] flux is low are once again impossible to analyze since the [N{\sc ii}]$\uplambda$5755 line is only detected in some bright knots around the close neighbourhood of the star. 

In Tables \ref{table2.4}, we present the relative fluxes and the extinction coefficient $c_{\text{H}\upbeta}$ for each of the seven regions indicated in Figure \ref{fig4}. Relative fluxes were reddening-corrected following to the technique described in Section \ref{section3.2}. In Table \ref{table2.5}, the electron temperature, density and ion abundances measured for these selected regions are presented and were computed with PyNeb. For the temperature derived from our regions (see Table \ref{table2.5}), most of the oxygen is in the form of O$^+$ and O$^{2+}$ and most of the nitrogen is in the form of N$^+$. Therefore we used the following approximations for the abundance ratios:
\begin{align}
    \rm \frac{O}{H} &= \rm \frac{O^+ + O^{2+}}{H^+}  \\
    \rm \frac{N}{O} &= \rm \frac{N^+}{O^+} \\
    \rm \log \frac{N}{H} &= \rm \log \frac{N}{O} + \log \frac{O}{H}
\end{align}
Unfortunately, although we detect faint O[{\sc iii}] emission (see Figure \ref{fig5}), our error bars are far too large to obtain reliable values of the flux in this line. Therefore, we adopt that $\log$(O/H) $\sim$ $\log$(O$^{+}$/H).

To properly evaluate the expected chemical abundance at the galactocentric radius of our nebula, we used the most recent distance estimate for WR124. \cite{2020MNRAS.493.1512R} recently used the parallax from the second Gaia Data Release \footnote{Gaia EDR3 was recently published and the observed parallax value was revised. The accepted value is now 0.15675 $\pm 0.01397$ mas, which lead to a distance of $6.38 \pm 0.57$ kpc.} to determine a distance for WR124 of $5.87^{+1.48}_{-1.09}$ kpc, which we will adopt in this study. Thus, we can assume that the galactocentric distance is between 6.56 and 6.76 kpc, which leads to an expected chemical abundance for nitrogen (\citet{2005ApJ...623..213C}) and oxygen (\citet{2005ApJ...618L..95E}) that would respectively be of 7.95 and 8.75. The values in Table \ref{table2.5} confirm that M1-67 is strongly enriched in nitrogen, by a factor of 4.3 on average and underabundant in oxygen by a factor of 4.9 on average. Nitrogen and oxygen abundance values from previous papers are similar and within the interval we determined in our seven regions (12 + $\log$(N/H) = [7.66:8.96] and 12 + $\log$(O/H) = [7.41:8.47] over the entire nebula). For comparison, \citetalias{1991A&A...244..205E} found values of  12 + $\log$(O/H) = 8.04 and 12 + $\log$(N/H) = 8.49, on average. These abundances are consistent with processed material from H-burning by the CNO cycle. 

From our limited dataset, we note that the lowest nitrogen enhancement is associated with the regions furthest from the star (for example, regions 5 and 7). However, the errors for these regions are not insignificant and don't allow us to reach a firm conclusion regarding the spatial variation in N enrichment. Nonetheless, these regions are close to the part of the nebula where the [N{\sc ii}]84/H$\upalpha$ ratio is highest ($\ge 3$). To explain such a pattern of high and low [N{\sc ii}]84/H$\upalpha$ ratio, an electron temperature variation of about $4\,000$ K between red and blue regions (Figure \ref{fig7}) is required. One could argue that region 5 displays the highest temperature from our sample ($9\,800 \pm 2\,500$ K), which corresponds to a variation of $3\,500$ K from the average temperature ($\sim 6\,300$ K). On the other hand, outer regions, such as regions 5 and 7, have higher uncertainties. Therefore, as we discussed earlier, it is premature to conclude that M1-67's characteristics can be explained by these extreme electron temperature variations. Although our temperatures are consistent with those from previous papers, they should not be overinterpreted.  In the next section, we present a kinematic analysis of the nebula in an attempt to explain this intriguing line-ratio behaviour. 

\section{Kinematics of M1-67} \label{section4}

The kinematics of M1-67 were first studied in detail by \citetalias{1998A&A...335.1029S}. From their high S/N long-slit spectroscopy, these authors fitted single or multiple Gaussian profiles to nebular emission lines for several bright clumps (over 13 positions) within the nebula. From the measured radial velocities, they suggested the presence of a hollow shell expanding with a velocity of 46\,km\,s$^{-1}$ with respect to a center of expansion of 137 km\,s$^{-1}$ and a separate bipolar outflow with a velocity of 88\,km\,s$^{-1}$. In a later study, \cite{2003A&A...398..181V} (hereafter SL03) took into account the fact that WR124 is a runaway star ($v_r \sim 190$\,km\,s$^{-1}$) and fitted the 3D Fabry-Perot data of \cite{GROSMOFF} to model the interaction of several outbursts from the star with the bow shock that results from the rapid movement of the star with respect to the ISM. With their model, they reproduced the appearance and kinematics of the nebula with various ejecta colliding with the bow shock shortly after the outbursts. The material ejected in front of the star is then dragged along the bow shock surface. As the star is moving away from us in nearly the direction of our line-of-sight (LOS), we observe a superposition of different contributions along the shock front. The gaz ejected between the star and the opening of the bow shock is, on the other hand, freely expanding. 

In order to understand the kinematics of the gas in M1-67, an important parameter is the heliocentric radial velocity of the WR star itself. It is widely accepted that the star is moving with a radial velocity of about +190 km/s (\citealt{2007AN....328..889K}). Although the star is saturated in our SN3 datacube, we were able to extract a spectrum from the unsaturated wings of the Point Spread Function and fitted simultaneously the broad stellar lines of He{\sc ii}/H$\alpha$ and He{\sc i}$\uplambda$6678. We obtained a heliocentric velocity of $185 \pm 9$ km/s, which is compatible with the value of \citet{2007AN....328..889K}.  Therefore, we will adopt this value for the remainder of our analysis for the velocity of WR124 itself.

\begin{landscape}
\begin{table}
    \centering
        \caption{Reddening-corrected line intensities (relative to H$\upbeta=100$) for each region as identified in Figure \ref{fig4}. The measured H$\upbeta$ surface brightness F(H$\upbeta$) is in erg cm$^{-2}$ s$^{-1}$ and the reddening coefficient c(H$\upbeta$) are also given. For each region, the number of summed spectra is different and is indicated at the bottom of this table. The last two rows give the offset from the star for each region in arcsec ($"$).}
    \renewcommand{\arraystretch}{1.5}  
    \scalebox{1.1}{
    \begin{tabular}{lccccccccc} \hline \hline
        \multicolumn{10}{c}{I($\uplambda$)/I$(\rm H\upbeta)$} \\ \hline
        Line & $\uplambda$ $(\si{\angstrom})$   & f($\uplambda$) & Region 1         & Region 2          & Region 3        & Region 4         & Region 5         & Region 6          & Region 7      \\ \hline
        \textbf{[O$_{\rm II}$]}    & 3727     & 0.322        & 22.6 $\pm$ 3.9   & 27.7 $\pm$ 7.4   & 15.9 $\pm$ 1.6   & 11.5 $\pm$ 1.0   & 48.2 $\pm$ 8.9   & 14.9 $\pm$ 1.5    & 26.6 $\pm$ 3.3    \\
        \textbf{H$\upbeta$}          & 4861     & 0.000        & 100.0            & 100.0            & 100.0            & 100.0            & 100.0            & 100.0             & 100.0           \\
        \textbf{[O$_{\rm III}$]}   & 5007     & -0.038       & 4.1 $\pm$ 1.8    & 4.2 $\pm$ 2.0    & 2.3 $\pm$ 2.0    & 3.0 $\pm$ 2.2    & 7.7 $\pm$ 2.5    & 2.5 $\pm$ 1.9     & 4.6 $\pm$ 2.1     \\
        \textbf{[N$_{\rm II}$]}    & 5755     & -0.185       & 0.8 $\pm$ 0.5    & 1.4 $\pm$ 1.8    & 0.8 $\pm$ 0.3    & 2.0 $\pm$ 0.4    & 7.8 $\pm$ 2.0    & 0.9 $\pm$ 0.2     & 2.2 $\pm$ 0.7     \\
        \textbf{He$_{\rm I}$}      & 5876     & -0.203       & 7.0$\pm$ 0.5     & 25.9 $\pm$ 1.7   & 2.4 $\pm$ 0.2    & 4.3 $\pm$ 0.4    & 7.1 $\pm$ 1.9    & 1.9 $\pm$ 0.2     & 5.1 $\pm$ 0.6     \\
        \textbf{[N$_{\rm II}$]}    & 6548     & -0.296       & 126.7 $\pm$ 2.4  & 137.2 $\pm$ 3.4  & 113.2 $\pm$ 2.8  & 112.0 $\pm$ 3.3  & 186.1 $\pm$ 3.5  & 111.8 $\pm$ 2.3   & 144.0 $\pm$ 2.7   \\
        \textbf{H$\upalpha$}         & 6563     & -0.298       & 325.2 $\pm$ 3.1 & 325.2 $\pm$ 4.2   & 327.2 $\pm$ 3.7  & 315.6 $\pm$ 4.9  & 324.2 $\pm$ 4.0  & 326.3 $\pm$ 3.1   & 325.4 $\pm$ 3.3 \\
        \textbf{[N$_{\rm II}$]}    & 6584     & -0.300       & 376.1 $\pm$ 3.3 & 406.1 $\pm$ 4.7   & 336.5 $\pm$ 3.8  & 338.3 $\pm$ 5.2  & 572.7 $\pm$ 5.2  & 334.8 $\pm$ 3.2   & 435.5 $\pm$ 3.8   \\
        \textbf{[S$_{\rm II}$]}    & 6717     & -0.318       & 13.1 $\pm$ 2.1   & 11.6 $\pm$ 3.0   & 12.2 $\pm$ 2.5   & 12.9 $\pm$ 2.8   & 22.0 $\pm$ 3.0   & 13.1 $\pm$ 2.1    & 16.2 $\pm$ 2.4    \\
        \textbf{[S$_{\rm II}$]}    & 6731     & -0.320       & 14.8 $\pm$ 2.1   & 15.6 $\pm$ 3.0   & 17.6 $\pm$ 2.6   & 19.2 $\pm$ 2.9   & 22.1 $\pm$ 3.0   & 19.2 $\pm$ 2.1    & 16.1 $\pm$ 2.4    \\
        \textbf{F(H$\upbeta$)}      & $10^{-14}$&  & 2.01 $\pm$ 0.04  & 0.48 $\pm$ 0.01  & 2.67 $\pm$ 0.06  & 5.56 $\pm$ 0.01  & 1.25 $\pm$ 0.03  & 3.56 $\pm$ 0.07   & 7.48 $\pm$ 0.16             \\    
        \textbf{c$_(\rm{H}\upbeta)$} &          &              & 1.41 $\pm$ 0.04  & 1.41 $\pm$ 0.05  & 1.55 $\pm$ 0.05  & 0.75 $\pm$ 0.03  & 1.34 $\pm$ 0.05  & 1.49 $\pm$ 0.04   & 1.43 $\pm$ 0.04 \\ \hline
        Nb of pixel    &          &              & 332              & 116              & 64               & 91               & 794              & 85                & 2 814                         \\
        $\Delta \upalpha$            &          &              & 22               & 9                & 0                & 0                & 38               & 9                 & 46              \\
        $\Delta \delta$            &          &              & 2                & 5                & 13               & 16               & 38               & 17                & 9      
    \end{tabular}
    }
    \label{table2.4}
\end{table}

\begin{table}
    \centering
        \caption{Electron densities (cm$^{-3}$), temperatures (K), ion abundances and total chemical abundances for the selected region.}
    \setlength{\tabcolsep}{8pt}    
    \renewcommand{\arraystretch}{1.3}
    \scalebox{1}{
    \begin{tabular}{lccccccc} \hline \hline
                                    & Region 1                  & Region 2        & Region 3          & Region 4          & Region 5          & Region 6          & Region 7        \\ \hline
        n$_e$([S$_{\rm II}$])       & 600 $\pm$ 200             & 1 200 $\pm$ 500 & 1 460 $\pm$ 510   & 1 800 $\pm$ 600   & 510 $\pm$ 140     & 1 500 $\pm$ 400   & 430$\pm$ 120  \\
        T$_e$([N$_{\rm II}$])       & 5 600 $\pm$ 850                    & 6 200:          & 5 800 $\pm$ 1 800 & 7 300 $\pm$ 1 500 & 9 800 $\pm$ 2 500 & 5 900 $\pm$ 1 300 & 7 000 $\pm$ 2 100 \\
        % 12 + $\log$(O$^+$/H$^+$)    & 8.47 $\pm$ 1.45           & 8.37 $\pm$ 2.23 & 8.44 $\pm$ 0.83   & 7.73 $\pm$ 0.70   & 7.41 $\pm$ 1.36   & 8.34  $\pm$ 0.83  & 7.66 $\pm$ 1.39 &   \\
        % 12 + $\log$(O$^{2+}$/H$^+$) & 6.79:           & 7.10:           & 6.79:           & 6.62:           & 6.45:           & 6.92:           & 6.73:           \\
        % 12 + $\log$(S$^+$/H$^+$)    & 6.66 $\pm$ 2.02           & 6.51 $\pm$ 2.96 & 6.72 $\pm$ 2.37   & 6.42 $\pm$ 2.34   & 6.07 $\pm$ 1.66   & 6.71 $\pm$ 1.83   & 6.39 $\pm$ 5.93 &   \\
        12 + $\log$(N$^+$/H$^+$)    & 8.72 $\pm$ 0.23           & 8.55 $\pm$ 0.30 & 8.62 $\pm$ 0.31   & 8.24 $\pm$ 0.34   & 8.07 $\pm$ 0.21   & 8.56 $\pm$ 0.25   & - \\
        $\log$(N$^+$/O$^+$)         & 0.39 $\pm$ 0.07           & 0.46 $\pm$ 0.14 & 0.52 $\pm$ 0.07   & 0.92 $\pm$ 0.11   & 0.77 $\pm$ 0.16   & 0.58 $\pm$ 0.07   & 0.83 $\pm$ 0.21 \\
        12 + $\log$(O/H)            & -   & - & - & -   & -   & -  & - \\
        % 12 + $\log$(N/H)            & 8.86 $\pm$ 1.45           & 8.83 $\pm$ 2.23 & 8.96 $\pm$ 0.83   & 8.65 $\pm$ 0.70   & 8.18 $\pm$ 1.36   & 8.92 $\pm$ 0.83   & 7.66 $\pm$ 1.39 &   \\
    \end{tabular}
    }
    \label{table2.5}
\end{table}
\end{landscape}

\subsection{One-component fits} \label{section4.1}

\begin{figure}
    \centering
    \includegraphics[width=\columnwidth]{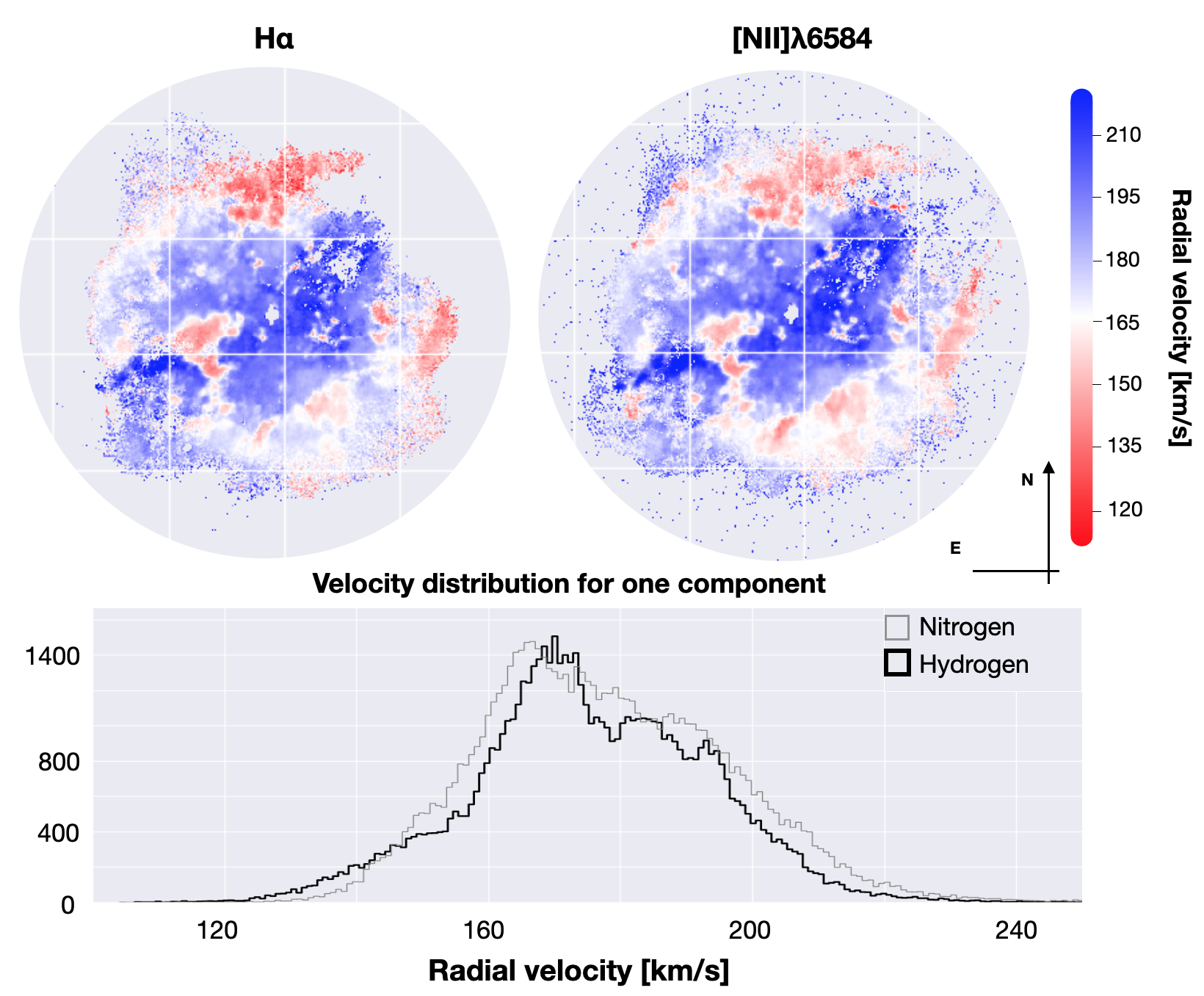}
    \caption{Velocity maps with one fitted component for the H$\upalpha$ and N[{\sc ii}]$\uplambda$6584 lines. Pixels displayed have a S/N$\ge$10. The associated histograms shown in the bottom panel shows the very similar velocity distributions for these two ions.}
    \label{fig10}
\end{figure}

In Figure \ref{fig10}, we present the velocity maps for the [N{\sc ii}]$\uplambda$6584 and H$\upalpha$ lines obtained from fitting one velocity component to our nebular lines. Note that for simplicity, we imposed the same velocity dispersion for both transitions. Both maps have an extremely similar appearance. In the bottom panel, we present the histogram of velocities from these maps, illustrating that the gas distributions for these two ions are nearly identical. 

\begin{figure}
    \centering
    \includegraphics[width=0.95\columnwidth]{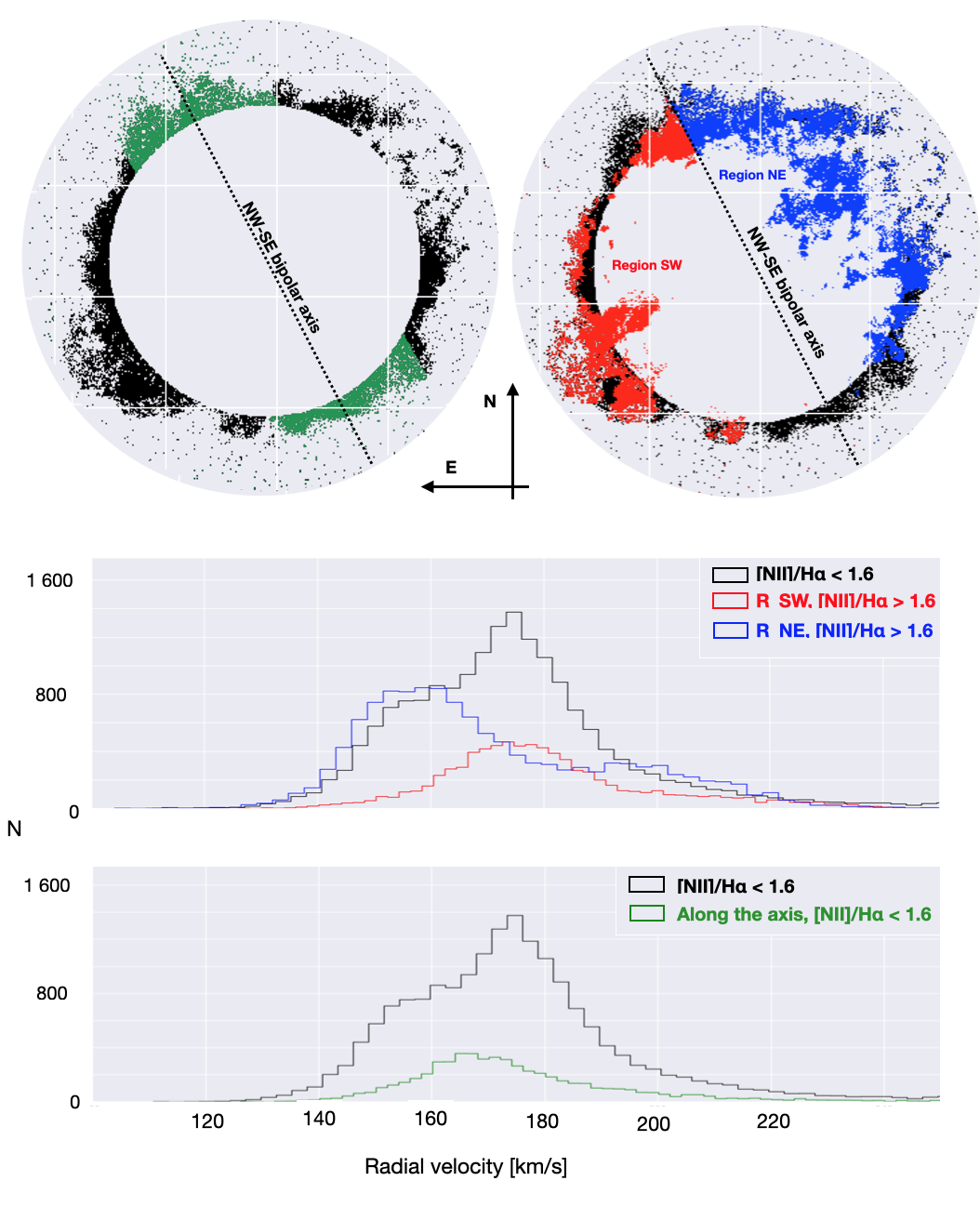}
    \caption{Velocity maps (top) and histograms (middle and bottom) of the [N{\sc ii}]$\lambda$6584 gas for regions where the [N{\sc ii}]/H$\upalpha$ line ration is greater than 1.6 (bleu and red) and smaller than 1.6 (green and black).  Pixels with [N{\sc ii}]/H$\upalpha$ $< 1.6$ and within a radial distance of $<42"$ of the central star were excluded from this analysis to avoid confusion in the central regions.}
    \label{fig11}
\end{figure}

As Figure \ref{fig7} clearly shows an intriguing behaviour of the [N{\sc ii}]/H$\upalpha$ line ratio along the NW-SE and NE-SW axes, we resolved to determine if this behaviour had any correspondance in the kinematics of the gas. Since we found no significant differences between the velocity histograms of the H$\upalpha$ and [N{\sc ii}] lines (see Figure \ref{fig10}), we analysed only the [N{\sc ii}]$\lambda$6584 velocities. First, because the middle of the nebula suffers from confusion due to the overlap of gas in the two directions for line ratio values smaller than 1.6, we kept only pixels further out than 42$"$ from the central star. This allowed us to isolate mainly the outer regions of the nebula, which constitute our main interest. Next, we attributed a colour code to all pixels based on their spatial location. We drew a line along the NE-SW axis, dividing the nebula in two. We also considerated the value of their [N{\sc ii}]/H$\upalpha$ line ratio. All pixels that had a line ratio smaller than 1.6 were coded in black. For the pixels with a ratio greater than 1.6, we refer to the NE-SW axis. All pixels to the west of this line were coded in blue and all pixels to the east in red. The final map is shown in the top right panel of figure \ref{fig11}. In the middle panel of this figure, we present the histograms of the velocities obtained with a one-component fit for the blue, red and black pixels. Finally, we added one final refinement. In order to verify if the gas with a [N{\sc ii}]/H$\upalpha$ line ratio smaller than 1.6 located along the NE-SW direction had a particular kinematic behaviour, we selected the pixels within a distance of 15$''$ of this axis with a ratio lower than 1.6 and coded them in green. The resulting map is shown in the top left panel of Figure \ref{fig11}. The bottom panel presents the histograms for the black and green pixels. Two conclusions can be drawn from these histograms. The first is that all pixels with a [N{\sc ii}]/H$\upalpha$ line ratio smaller than 1.6 have a similar velocity distribution to the velocity distribution shown in Figure \ref{fig10}. The final conclusion is that pixels with a ratio larger than 1.6 have a different velocity distribution depending on their location in the nebula. Pixels in the NW part of the nebula (coded in blue) have a velocity distribution centered at $\sim$160\,km\,s$^{-1}$ and those in the SE part of the nebula (coded in red)  at $\sim$185\,km\,s$^{-1}$.  Such a behaviour could be explained if we were viewing a bipolar ejection nearly in the plane of the sky but slightly tilted with respect to our line-of-sight.

% \citet{thesethomas} present 

\subsection{Two-components spectral fits} \label{section4.2}

In spite of the moderate spectral resolution of our observations, the ORCS software can be used to carry out fits with two velocity components for strong nebular lines, as the line shapes remain slightly affected if two components are present along the line of sight. In addition to the velocity of the gas, the line fits with one velocity component presented in the previous section provides us with a map of line widths, corresponding to the $\sigma$ parameter describing the Gaussian function convolved with the sincgauss. Our velocity dispersion map is presented in Figure \ref{fig9}. We then used the value of $\sigma$ to determine if, at a given position, one or two velocity components were more appropriate to represent our data. For a given position in our maps, we fitted one of the following :

\begin{figure}
    \includegraphics[width=\columnwidth]{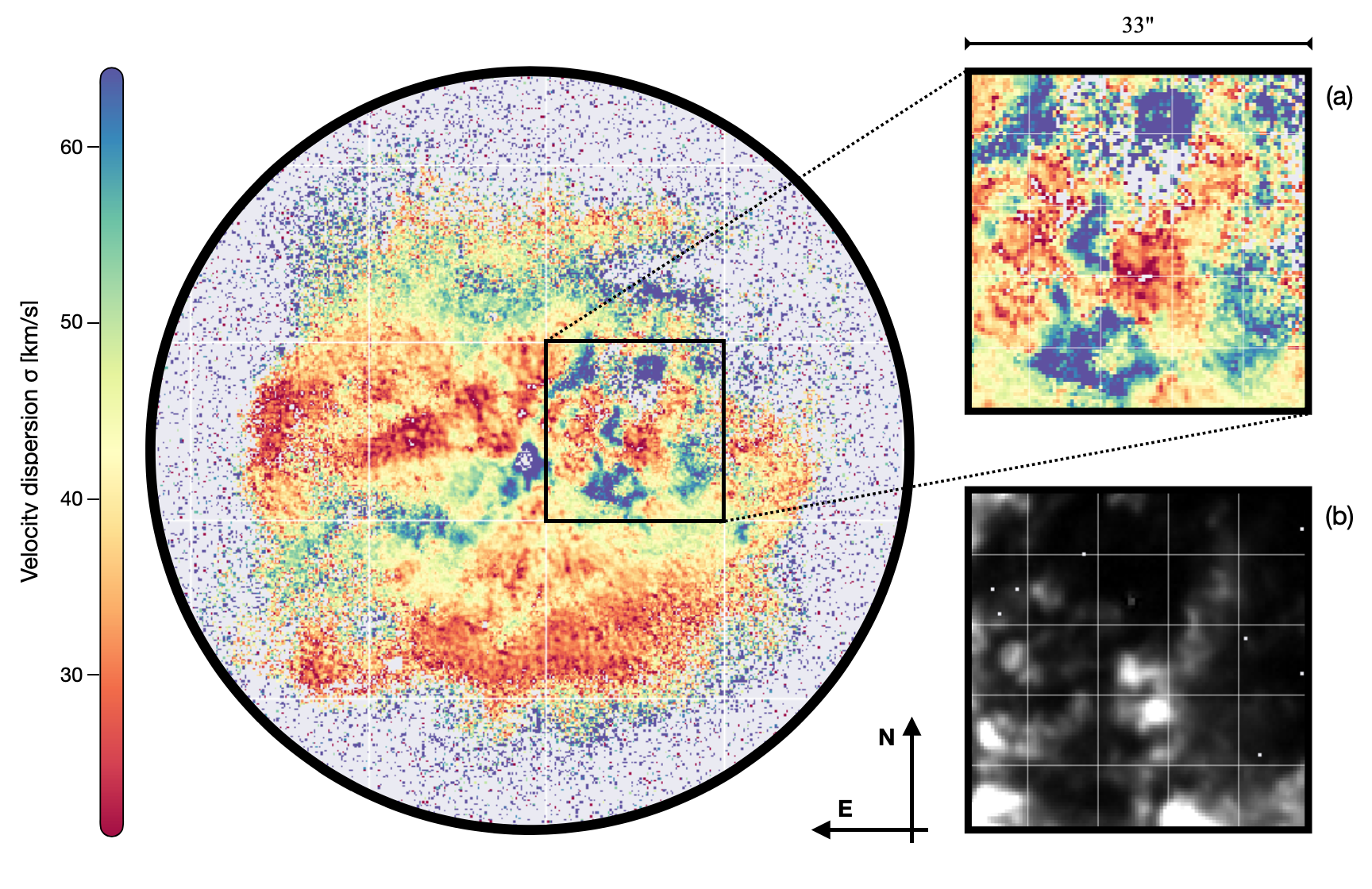}
    \caption{Velocity dispersion map of M1-67 for the H$\upalpha$ line from the SN3 filter. In panel (a), we selected a region that includes a large range in velocity dispersion (from 30 to 60 km/s) and numerous individual structures. To facilitate direct comparison, we present an H$\upalpha$ flux map of the same region in panel (b).}
    \label{fig9}
\end{figure}

\begin{itemize}
    \item A single sincgauss function. This profile is adequate when the broadening is small compared to the channel width. This is described in detail in \cite{2016MNRAS.463.4223M}. For this case, the width of the emission line and its velocity are the only two parameters that characterize the emission line.
    \item A two components sinc model where each emission line is modelled as the sum of two resolved sinc functions. This case is better adapted for larger line widths ($\upsigma$). This case, of course, requires twice the number of fitting parameters but for simplicity, we used the same width for both components and we excluded every position where the error on the velocity dispersion was greater than 5\,km\,s$^{-1}$.
\end{itemize}

The threshold between these two cases was for a value of 30 km s${^{-1}}$ and was determined as follows.  We carried out two-component fits to lines of different widths using the ORCS code. The result was that below this value, the resulting fits depended greatly on our initial velocity guesses for each component. For widths larger than 30 km s${^{-1}}$ on the other hand, the two-component fits converged to the same solution no matter which initial parameters we used, indicating the robustness of the method for lines with widths above this threshold.

\begin{figure}
    \includegraphics[width=\columnwidth]{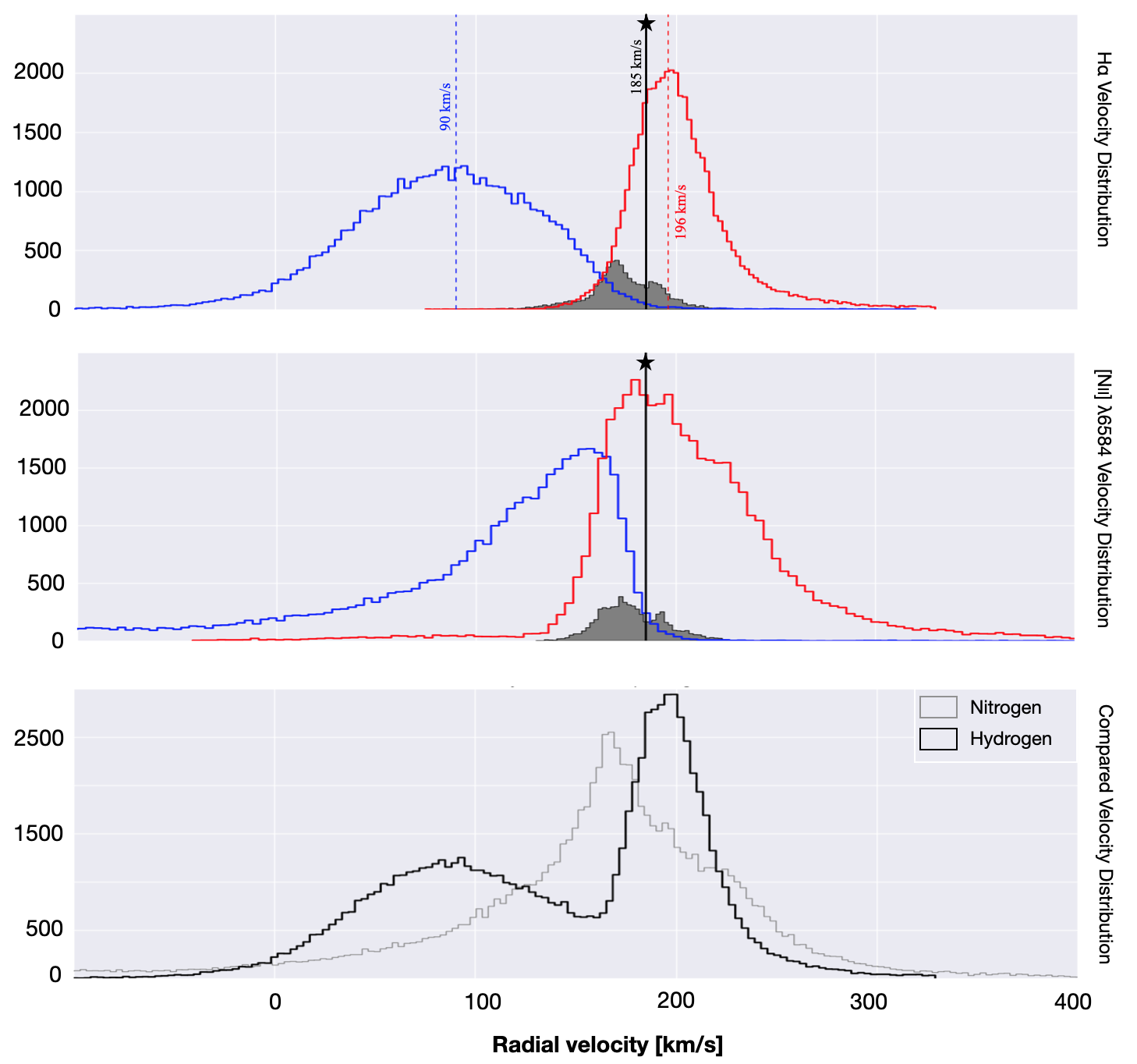}
    \caption{Heliocentric velocity distribution for all pixels with an S/N ratio $\ge 3$ for the hydrogen (top) and nitrogen (middle) gas. For the top and middle panels, the red and blue distributions represent the two fitted components while the black histogram is for lines with a width smaller than 30 km s${^{-1}}$. The black vertical line indicates our measured heliocentric velocity of the central star ($\sim$185\,km\,s$^{-1}$). The third panel compares the combined velocity distributions of both elements.}
    \label{fig12}
\end{figure}

 The histograms of the heliocentric radial velocities for the H$\upalpha$ and [N{\sc ii}]$\uplambda6584$ lines for the two-component fits are presented in Figure \ref{fig12}. In the top and middle panels, we present the velocity distributions for the hydrogen and nitrogen gas respectively. In black, we plot the velocities obtained for the lines with widths below 30 km s${^{-1}}$ and in blue and red, the velocities of both fitted components.  The bottom panel compares the velocity distribution of the hydrogen and nitrogen gas when we combine the two fitted components. The black vertical line indicates our measured heliocentric velocity for the central star ($\sim$185\,km\,s$^{-1}$). Contrary to the results for the single-component fits, the distributions for H$\upalpha$ and [N{\sc ii}]$\uplambda$6584 lines in this case are quite different. 
 
 For H$\upalpha$, we distinguish two well-separated distributions, one centered at $\sim$+195 km\,s$^{-1}$ that is relatively narrow and the other at $\sim$+90 km\,s$^{-1}$ that is much broader. For the [N{\sc ii}]$\uplambda$6584 line, we also distinguish two components but they are much closer together. One is centered near $\sim$+200 km\,s$^{-1}$ and the other near $\sim+$150 km\,s$^{-1}$. The difference between the [N{\sc ii}] and H$\upalpha$ are clear in the bottom panel where we compare the combined distributions for these two ions.
 
\begin{figure}
    \centering
    \includegraphics[width=1\columnwidth]{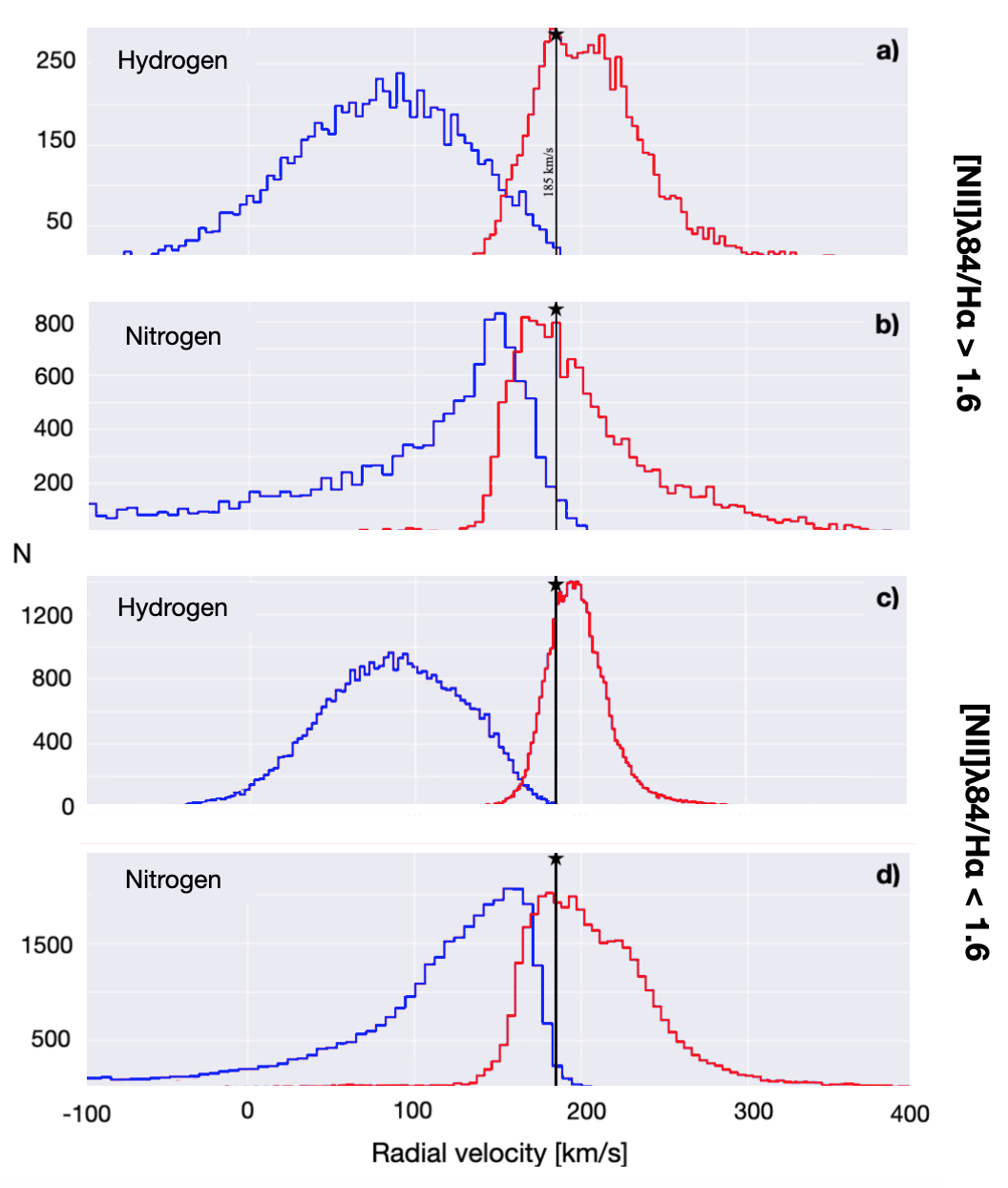}
    \caption{Radial velocity distribution for the two-components fits for pixel with an S/N ratio $\ge 3$ for the nitrogen and hydrogen gas. The first two top panels are for regions with a [N{\sc ii}]$\uplambda$6584 larger than 1.6 (see Fig. \ref{fig7})} and the bottom two panels for when this ratio is lower than 1.6.
    \label{fig13}
\end{figure}

 In the different panels of Figure \ref{fig13}, we present separate two-component-fits velocity distributions for gas with [N{\sc ii}]84/H$\upalpha$ higher (top two panels) and lower (bottom two panels) than 1.6. The positions of the associated points in the nebula can be determined from Figure \ref{fig7}. We were not able to confirm the bipolar ejection suggested by \citetalias{1998A&A...335.1029S}. In fact, contrary to what we found for our one-component fits, we find no significant differences in the velocity distributions. In this case, this line ratio does not seem to be a determining factor in the kinematics of the gas. 

\subsection{Bow-shock kinematics} \label{section4.3}

In an attempt to shed some light on the kinematics of the gas in M1-67, we used our two-component fits to verify if our measured velocities are compatible with the kinematics of the bow-shock model introduced by \citetalias{2003A&A...398..181V}.  These authors present the first study of the dynamics of ejected gas interacting with the bowshock formed as the central star moves at high velocity through the interstellar medium. Using H$\alpha$ Fabry-P\'erot observations from \citet{1999IAUS..193..356G} they were able to determine the orientation of this paraboloid bowshock with respect to our line-of-sight. To characterize their models, \citetalias{2003A&A...398..181V} used three main parameters:  the velocity of the star with respect to the ISM ($v_{\rm ism}$) and two orientation parameters, the inclination of the bowshock axis with respect to our line of sight, $i$ and the angle in the plane of the sky between the shockcone axis and our line of sight, $\upphi$. Their fits yield values of $i$=20$^o$, $\upphi$=$-$185$^o$ and $v_{\rm ism}$=180 km\,s$^{-1}$ meaning that our line-of-sight is nearly aligned with the axis of the bow shock, looking into the opening of the cone. They find evidence for two distinct outbursts having occurred at different times inside the bow shock. Their model shows that these outflows have a velocity of $\sim$150 km\,s$^{-1}$.
 
\begin{figure*}
    \centering
    \includegraphics[width=0.8\textwidth]{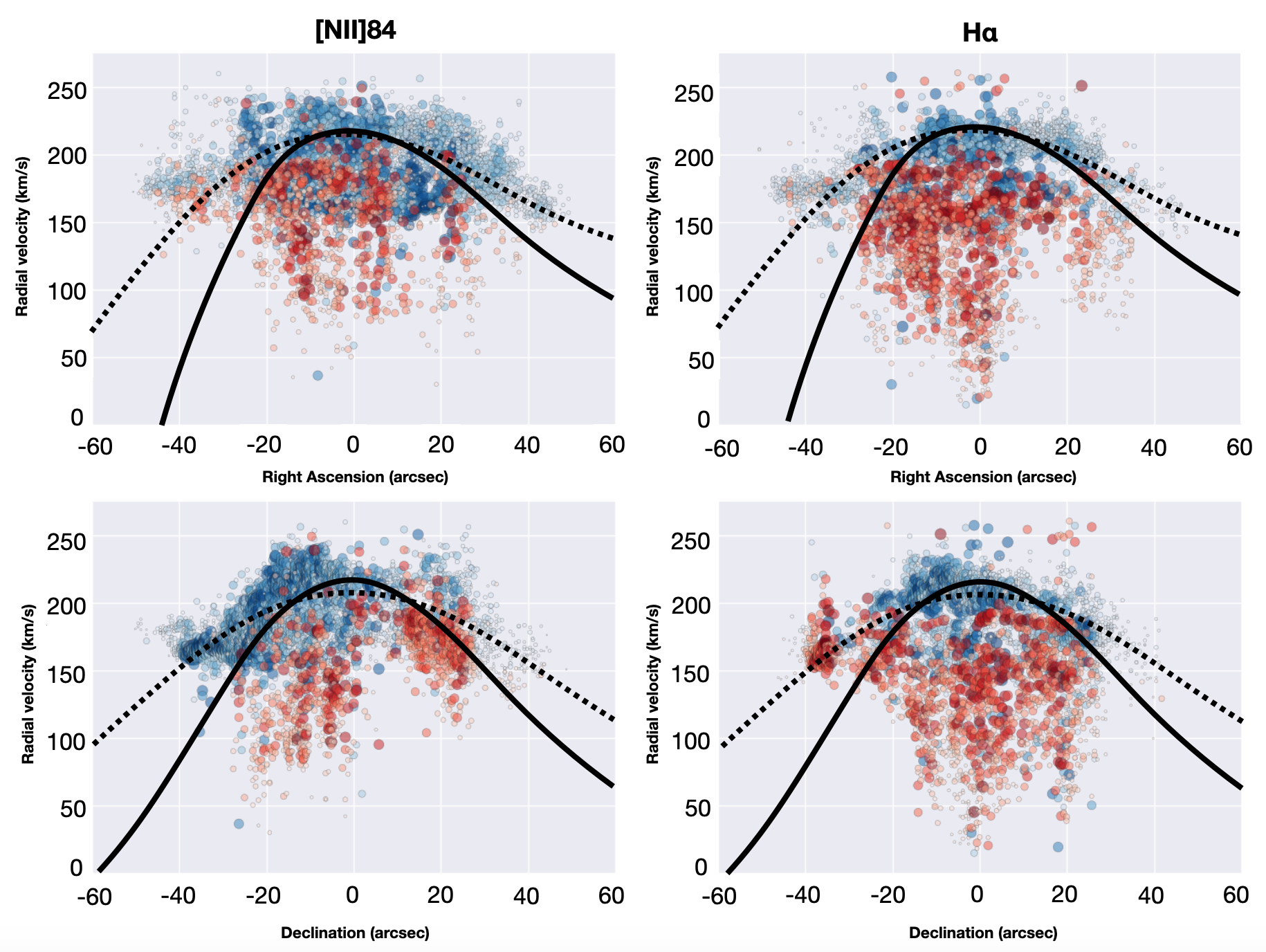}
    \caption{Upper panels : radial velocity against right ascension. Lower panels : radial velocity against declination where both components of the resolution are shown with the same colour code used in Figure \ref{fig12} and \ref{fig13}. The size and colour of each point are proportional to the flux measured in the [N{\sc ii}]84 line for the first column and for the H$\upalpha$ line for the second column. The dashed and solid lines are model outputs for the best 3D bow shock model for WR124 from \citetalias{2003A&A...398..181V} with a distance d$=6.0$ kpc and d$=8.0$ kpc respectively.}
    \label{fig14}
\end{figure*}

Figure \ref{fig14} presents our measured velocities in the right ascension-radial velocity and declination-radial velocity plane for both the [N{\sc ii}]$\uplambda$6584 and H$\alpha$ lines. 
these plots are very similar to the one presented by \citetalias{2003A&A...398..181V} except that in their case, they had a lack of observations at declination offset larger than -20$''$ because of bad seing. Superposed on our observations are model outputs from their 3D bowshock model for distances of 6.0 kpc and 8.0 kpc (respectively dashed and solid lines), a total velocity of the star with respect to the iSM of $v_{\rm ism} = 180$ km\,s$^{-1}$ and an orientation of the shock cone of $i=20^{\circ}$ and $\upphi = 185^{\circ}$. It is clear that the model using the recently determined Gaia distance fits well with our observations. We also can see the material ejected and freely expanding inside the bow shock, probably causing the spread and slow component in the Figure \ref{fig12}.

\section{Conclusions} \label{Conclusion}

M1-67 is a "long-time" challenging and well-known object. Our data provides a few new pieces of evidence to help understanding its origins :

(i) The [NII]/H$\upalpha$ ratio map (Figure \ref{fig7}) clearly reveals a patent asymetry in the outer regions of the nebula. Although the entire nebula is enriched in nitrogen, the outer SE and NW regions, filamentary and clumpy, are characterized by an even stronger nitrogen enhancement: the [NII]/H$\upalpha$ ratio can reach 5 - 10. We also note the presence of an extension of these structures to the north. To the contrary, the faint, very diffuse structures at the NE and SW corners display anomalously low [NII]/H$\upalpha$ ratios compared to the rest of the nebula. These four regions are characterized by almost identical average radial velocities (170 $\pm$ 1.5 km/s). Although drastically different electron temperatures could reproduce these changing line ratios, these very likely represent truly different nitrogen enhancements. Getting the information about the rotation axis of the star could help us to interpret the orthogonal and elongated pattern with anisotropic winds or ejections. Such a link between the rotation axis and enhancement axis could point in the direction of the anisotropic winds caused by high velocity rotation.

(ii) A few thin radial filamentary structures clearly stand out in the [NII]/H$\upalpha$ map, with a significantly lower [NII]/H$\upalpha$ ratio than their immediate surroundings. These are located on the opposite side of dense H$\upalpha$ knots, as seen from the star. The exact origin of those structures is still unknown.

(iii) In addition to the strong lines, we also detect the presence of the fainter [O{\sc iii}]$\uplambda$ 5007, [O{\sc ii}]$\uplambda$ 3727-29, He{\sc i} $\uplambda$6678, He{\sc i} $\uplambda$5876 and [N{\sc ii}]$\uplambda$ 5755 in some integrated regions, allowing us to estimate electronic temperature and chemical abundance.

(iv) Our one component fits revealed a different velocity distribution for the gas with [N{\sc III}]/H$\upalpha$ larger or smaller than 1.6.

(v) We also performed two-components fits on our spectra and found our radial velocity measurements to be in good agreement with those of previous studies, particularly with the 3D bow-shock model of \citetalias{2003A&A...398..181V} .  The model corresponding to a distance d$=6.0$ kpc was found to best correspond to our observed velocity distributions, which agrees well with the determination of the distance of the nebula by \citet{2020MNRAS.493.1512R} using recent Gaia observations.

(vi) From a kinematic perspective, we confirm the role of the bow shock formed by this fast moving stars on the velocity of the gas in M1-67. Our velocity maps can be interpreted in the context of a model in which this ejected nebula is the result of many spherical and non-spherical outbursts interacting with the bow shock structure. 

These results clearly demonstrate how SITELLE is well adapted for the study of such emission-line nebulae surrounding massive stars in late evolutionary phases such as WR124.

\section*{Acknowledgements}
Based  on  observations  obtained  with  SITELLE,  a  joint project  of  Universit\'e  Laval,  ABB,  Universit\'e  de  Montr\'eal and  the  Canada-France-Hawaii  Telescope  (CFHT). which is  operated  by  the  National  Research  Council of Canada,  the  Institut  National  des  Sciences  de  l'Univers  of the  Centre  National  de  la  Recherche  Scientifique of France and the University of Hawaii. The authors wish to recognize and acknowledge the very significant cultural role that the summit of Mauna Kea has always had within
the indigenous Hawaiian community. We are most grateful to have
the opportunity to conduct observations from this mountain.
LD and are grateful to the Natural Sciences and Engineering Research Council of
Canada, the Fonds de Recherche du Qu\'ebec, 
and the Canada Foundation for Innovation for financial support.

\section*{Data Availability}
The data underlying this article will be shared on reasonable request to the corresponding author.

%%%%%%%%%%%%%%%%%%%% REFERENCES %%%%%%%%%%%%%%%%%%

% The best way to enter references is to use BibTeX:

\bibliographystyle{mnras}
\bibliography{bibliography} % if your bibtex file is called example.bib

% Don't change these lines
\bsp    % typesetting comment
\label{lastpage}
\end{document}